\documentclass[%
 reprint,
superscriptaddress,
 amsmath,amssymb,
 aps,
 pra,
floatfix,
]{revtex4-1}

\usepackage{amsfonts}
\usepackage{amsmath}
\usepackage{amsthm}
\usepackage{mathrsfs}
\usepackage{amscd}
\usepackage{amssymb}
\usepackage{subfigure}
\usepackage{amsxtra}
\usepackage{dcolumn}% Align table columns on decimal point
\usepackage{bm}           % bold math
\usepackage{bbm}
\usepackage{graphicx}

\usepackage[colorlinks]{hyperref}
\newcommand{\all}[1]{\underline{#1}}

\renewcommand{\vr}{\boldsymbol{r}}
\newcommand{\ve}{\boldsymbol{e}}
\newcommand{\vx}{\boldsymbol{x}}
\newcommand{\va}{\boldsymbol{a}}
\newcommand{\vb}{\boldsymbol{b}}
\newcommand{\vz}{\boldsymbol{z}}
\newcommand{\vZ}{\boldsymbol{Z}}

\newcommand{\vk}{\boldsymbol{k}}
\newcommand{\vF}{\boldsymbol{F}}
\newcommand{\vR}{\boldsymbol{R}}

\newcommand{\vp}{\boldsymbol{p}}
\newcommand{\vq}{\boldsymbol{q}}

\newcommand{\la}{\langle}
\newcommand{\ra}{\rangle}
\newcommand{\qerw}[1]{\ensuremath{\left\langle #1 \right\rangle}}

\newcommand{\cL}{\mathcal{L}}

\newcommand{\da}{\dagger}

\newcommand{\Op}[1]{\hat{#1}}

\newcommand{\oL}{\Op{L}}

\newcommand{\oU}{\Op{U}}
\newcommand{\ovr}{\Op{\vr}}
\newcommand{\ovx}{\Op{\vx}}
\newcommand{\ovp}{\Op{\vp}}
\newcommand{\ovR}{\Op{\vR}}

\newcommand{\op}{\Op{p}}
\newcommand{\ox}{\Op{r}} %cartesian components of r

\newcommand{\Mat}[1]{\mathbbm{#1}}

\newcommand{\id}{\ensuremath{\mathbbm 1}}
\newcommand{\tr}{\ensuremath{{\rm tr}}}
\newcommand{\erf}{\ensuremath{{\rm erf}}}
\newcommand{\sinc}{\ensuremath{{\rm sinc}}}

\newcommand{\diff}{\mathrm{d}}

\begin{document}

%\preprint{APS/123-QED}
 
\title{Classical Channel Gravity in the Newtonian Limit}
\author{Kiran E. Khosla}
\email{k.e.khosla@swansea.ac.uk}
\affiliation{Department of Physics, Swansea University, Swansea SA2 8PQ, United Kingdom}
\affiliation{Center for Engineered Quantum Systems, University of Queensland,
St Lucia 4072, Australia}

\author{Stefan Nimmrichter}
\email{cqtsn@nus.edu.sg}
\affiliation{Centre for Quantum Technologies, National University of Singapore, 3 Science Drive 2, Singapore 117543, Singapore}

\date{\today}% It is always \today, today,

\begin{abstract}
We present a minimal model for the quantum evolution of matter under the influence of classical gravity in the Newtonian limit. Based on a continuous measurement-feedback channel that acts simultaneously on all constituent masses of a given quantum system, the model scales and applies consistently to arbitrary mass densities, and it recovers the classical Newton force between macroscopic masses. The concomitant loss of coherence is set by a model parameter, does not depend on mass, and can thus be confined to unobservable time scales for micro- and macroscopic systems alike. The model can be probed in high-precision matter-wave interferometry, and ultimately tested in recently proposed optomechanical quantum gravity experiments.
\end{abstract}
\maketitle     

\section{Introduction}\label{sec:intro}

Whether gravity is a quantum interaction or a fundamentally classical one is a question yet to be answered. Ultimately, this question must be answered experimentally, and recent proposals \cite{Pikovski2012,Pfister2016,Marletto2017,Bose2017,Carney2018} for an optomechanical test of the quantum nature of gravity have raised the hopes that a conclusive answer is within reach.

On the theory side, alternative hypotheses contrasting the well-established research program on quantum gravity \cite{Rovelli2004,Oriti2009,Kiefer2012} have garnered attention, according to which gravity might be a purely classical interaction channel after all, or embody a process inducing classical behavior at macroscopic scales. This includes models of state collapse by self-gravity \cite{Karolyhazy1966,Diosi1987,Penrose1996,Gasbarri2017a} and mass-induced spontaneous collapse \cite{Ghirardi1990a,Ghirardi1990b,Bassi2003,Bassi2012,Bassi2017}, as well as the postulate of a nonlinear Schr\"{o}dinger-Newton equation \cite{Ruffini1969,Diosi1984,Carlip2008,Giulini2011,Giulini2012,Bahrami2014d} and stochastic modifications of it \cite{Bera2015,Singh2015,Nimmrichter2015,Bera2017}. More recent works have shifted the focus from the collapse-inducing nature of a classical gravity to the emergence of an effective force between quantum mass densities. Kafri, Taylor, and Milburn (KTM) describe gravity as a classical communication channel, and they propose a measurement-feedback master equation  for an approximately linear force between test masses, and for discretized space \cite{Kafri2014,Kafri2015}. Tilloy and Di\'{o}si show how to incorporate an effective Newton potential into the known collapse models \cite{Tilloy2016,Tilloy2018}.

Here we present a minimal model for classical channel gravity (CCG) that is valid for arbitrary mass distributions in the Newtonian limit. It can be cast into a Galilei-covariant and exchange-symmetric master equation that models gravity as a multiplexed measurement-feedback process acting on all test masses in a given system. The model generalizes the idea of KTM: It recovers the desired classical limit of an average $1/r^2$-force between macroscopic test masses at macroscopic distances (and a regularized force at short distances). At the same time, it predicts the decay of coherence at a fixed rate that does \emph{not} amplify with mass---in stark contrast to the common feature of collapse models aiming to reinstate macroscopic realism \cite{Leggett2002,Nimmrichter2013}. As such, the present model would serve as a natural counter-hypothesis to test quantum gravity against in future experiments. 
%We note that, contrary to Diosi's gravitational collapse and similar models [CITE], CCG does not impose gravitational self-interaction of elementary test masses and is thus not plagued by the problem of diverging heating rates for point masses [CITE]. 

\section{Model}\label{sec:model}

CCG postulates that the gravitational interaction between test masses is mediated by an exchange of classical information rather than quantum information. This implies, in particular, that gravity cannot generate entanglement \cite{Marletto2017,Bose2017} and that it cannot be described \emph{solely} by means of an interaction potential in the system Hamiltonian. 
If we assume, at the same time, that the linear Hilbert-space framework of quantum theory remains intact, then gravity can be understood as a LOCC protocol \cite{Kafri2014} comprised of three steps: locate the the test particles, send these locations to all other particles, apply local unitaries to each particles that establish the net gravitation towards the other masses.

In the Newtonian limit of nonrelativistic gravity without retardation effects, the CCG protocol can be accommodated by a continuous measurement-feedback master equation \cite{Wiseman2010,Jacobs2014}. Consider a system of $N$ particles with masses $\all{m} := m_1,\ldots, m_N$ and (using the short-hand notation from now on) position and momentum operators $\all{\ovr}, \all{\ovp}$. The simultaneous monitoring of the particle positions is described by a product of $N$ weak position measurements,
\begin{equation}\label{eq:measN}
\mathcal{M} (\all{\ovr}-\all{\vz}) = \bigotimes_{n=1}^N \frac{\exp \left[ - (\ovr_n - \vz_n)^2/4\sigma_n^2 \right]}{\left( 2\pi \sigma_n^2 \right)^{3/4}},
\end{equation}
where the standard deviations $\sigma_n$ determine the measurement resolution per particle, and  $\int \diff^{3N} \all{z} \,\mathcal{M}^2 (\all{\ovr}-\all{\vz}) = \id$. Conditioned on the obtained measurement record $\all{\vz}$, each particle is subjected to a unitary transformation, $\oU(\all{\vz}) = \bigotimes_n \oU_n (\all{\vz}) $, which immediately follows the measurement and generates an effective gravitational attraction between the masses. Given the monitoring rate $\gamma$, we arrive at the master equation
\begin{equation} \label{eq:CCG_ME}
\cL_N \rho = \gamma \left[ \int\diff^{3N} \all{z} \, \oU (\all{\vz}) \mathcal{M} (\all{\ovr}-\all{\vz}) \rho \mathcal{M} (\all{\ovr}-\all{\vz}) \oU^\da (\all{\vz}) - \rho \right],
\end{equation}
which is added to the von Neumann equation of arbitrary interacting (excluding gravity) or non-interacting $N$-particle systems. 
In order for this master equation to represent the effect of non-relativistic classical gravity in a universal, consistent, and minimally invasive manner, it should: (i) add a Newtonian force term to the average equations of motion $\diff \la \ovp_n \ra / \diff t$ that predicts the correct classical behavior (e.g.~acceleration in earth's gravity, Kepler orbits), (ii) preserve Galilei covariance, (iii) preserve the exchange symmetry of identical quantum particles, (iv) ensure a consistent mass scaling and treatment of composite particles, and (v) cause minimal disturbance (e.g.~heating and decoherence). 
The requirements are met if we use
\begin{eqnarray} \label{eq:CCG_U}
\oU (\all{\vz}) &=& \bigotimes_{n=1}^N \exp \left[ \frac{i}{\hbar} \vq_n (\all{\vz}) \cdot \ovr_n \right],  \\
\vq_n (\all{\vz}) &=& -\frac{1}{\gamma} \sum_{k\neq n} \nabla \Phi_{nk} (|\vz_n - \vz_k|), \quad \sigma_n^2 = \frac{m_0}{m_n} \sigma_0^2 . \nonumber
\end{eqnarray}
The unitaries describe a momentum displacement on each particle according to an average gradient force $\gamma \vq_n (\all{\vz})$ conditioned on the measurement record. The factor of $\gamma$ ensures the correct scaling of displacements between monitoring events. If the associated two-particle potentials are chosen symmetric, $\Phi_{nk} (z) = \Phi_{kn} (z)$ (as required for Newtonian action-reaction pairs), Galilei covariance follows as $\sum_n \vq_n (\all{\vz}) = 0$, i.e.~there is no net center-of-mass force. In addition, the CCG master equation \eqref{eq:CCG_ME} is symmetric under the exchange of identical particles, as one can show by substitution of variables. 

Similar to a broad class of macrorealistic collapse models \cite{Adler2004,Bassi2012,Nimmrichter2013}, the CCG model is fully specified by two free parameters: a rate $\gamma$ at which the particles are interrogated, and a reference value $\sigma_0$ for the measurement resolution at an arbitrarily chosen reference mass $m_0$.

For a single isolated particle, there is no conditional unitary displacement, and \eqref{eq:CCG_ME} reduces to a standard Gaussian decoherence channel similar to that of Ghirardi, Rimini, and Weber~\cite{Ghirardi1986,Vacchini2007b},
\begin{eqnarray}
\cL_1 \rho &=& \gamma \left[ \int \frac{\diff^3 z}{(2\pi\sigma^2)^{3/2}} e^{-(\ovr-\vz)^2/4\sigma^2} \rho e^{-(\ovr-\vz)^2/4\sigma^2} - \rho \right] \nonumber \\
 &=& \gamma \left[ \int \frac{\sigma^3 \diff^3 k}{(\pi/2)^{3/2}} e^{-2\sigma^2 k^2} e^{i\vk \cdot \ovr} \rho e^{-i\vk \cdot \ovr} - \rho \right], \label{eq:CCG_ME1}
\end{eqnarray}
with $\sigma = \sqrt{m_0/m}\sigma_0$. It results in spatial decoherence with a rate of at most $\gamma$ and isotropic heating of the kinetic energy at the power $\mathcal{P} = 3\gamma \hbar^2/8m_0\sigma_0^2$. 
The same result is obtained for the reduced state of $n=1$ out of $N$ particles, which is so far away from the others that their interaction terms $\Phi_{1k}$ in \eqref{eq:CCG_U} can safely be neglected.

The same single-particle treatment applies also the center of mass coordinate---and thus universally to any composite test mass. 
Indeed, we show in Appendix \ref{app:COM} that tracing over the $N-1$ relative coordinates in \eqref{eq:CCG_ME} results in the single-particle form \eqref{eq:CCG_ME1} for the reduced center of mass state, $ \tr_{\rm rel} \{\cL_N \rho \} = \cL_1 \tr_{\rm rel} \{ \rho \}$, with the standard deviation $\sigma = \sqrt{m_0/M} \sigma_0$ corresponding to the total mass $M = \sum_n m_n$. This mass scaling is based on a simple metrological argument: If the position of $N$ identical test masses $m_0$ is simultaneously probed with an uncertainty $\sigma_0$, then the $N$ results, through their arithmetic mean, can be seen as $N$ independent measurements of the center-of-mass coordinate. Hence the uncertainty is reduced by $1/\sqrt{N}$.

The mass dependence has notable consequences for the macrorealistic nature of CCG. 
Most collapse models predict a decay of coherence that amplifies (at most quadratically) with mass -- a clear incentive to pursue high-mass interferometry to test them. 
Here, the CCG decoherence rate for systems of \emph{any} mass is always at most $\gamma$, and there is no intrinsic mass amplification. Isolated particles heat up at the (particle-mass independent) rate $\mathcal{P}$, and the heating may only enhance through the gravitational interaction with other surrounding masses.
This precludes experimental test schemes based on the center-of-mass heating or decoherence of isolated macroscopic masses, at least for all but the most invasive CCG parameter values $\gamma,\sigma_0$.

\emph{Force and diffusion.---}
The CCG momentum displacements \eqref{eq:CCG_U} are conditioned on the results $\all{\vz}$ and are applied immediately after each weak position measurement. They induce the average force $\partial_{t,{\rm CCG}} \la \ovp_n \ra = \la \vF_n (\all{\ovr}) \ra $ on each particle, where
\begin{equation} \label{eq:CCGforce1}
\vF_n (\all{\hat{\vr}}) = - \nabla_{\hat{\vr}_n} \sum_{k\neq n} \int \frac{ e^{-z^2/2\sigma_{nk}^2} \diff^3 z}{(2\pi \sigma_{nk}^2)^{3/2}}  \Phi_{nk} (|\hat{\vr}_{nk} - \vz|)
\end{equation}
is a Gaussian-smeared sum of the two-body terms $\Phi_{nk} (|\hat{\vr}_{nk}|)$, with $\hat{\vr}_{nk} = \hat{\vr}_n - \hat{\vr}_k$.  
The finite resolution $\sigma_0$ of the CCG protocol leads to a short-distance regularization of the gravitational forces acting on each particle, even for unregularized Newton potentials $\Phi_{nk}(r) = -Gm_nm_k/r_{nk}$ between point masses. This effect is also present when gravity is incorporated into collapse models \cite{Tilloy2016,Tilloy2018}. 
The resulting force \eqref{eq:CCGforce1} on the test mass $m_n$ would be that of the gravitational field sourced by $N-1$ Gaussian densities of masses $m_k$, and with standard deviations $\sigma_{nk} = \sqrt{m_0/\mu_{nk}} \sigma_0$ that depend on the relative masses $\mu_{nk} = m_n m_k/(m_n+m_k)$.

However, there is still room for divergences since the gravitational displacements \eqref{eq:CCG_U} also contribute to the CCG-induced momentum diffusion. In Appendix \ref{app:forceDiff}, we find that the second moments of momenta evolve as
\begin{eqnarray}
\partial_{t,{\rm CCG}} \left\la \ovp_n \ovp_k \right\ra &=& \hbar^2 \gamma \left[ \frac{\delta_{nk}}{4\sigma_n^2}\id + \qerw{\Mat{A}_{nk} (\all{\ovr}) } \right] \label{eq:CCG_2ndMoments} \\
&& + \frac{1}{2} \qerw{ \{ \ovp_n, \vF_k (\all{\ovr}) \} + \{ \vF_n (\all{\ovr}), \ovp_k \}}, \nonumber \\
\Mat{A}_{nk} (\all{\vr}) &=& \int \diff^{3N} \all{z} \, \mathcal{M}^2 (\all{\vr}-\all{\vz}) \frac{\vq_n (\all{\vz}) \vq_k (\all{\vz})}{\hbar^2}, \label{eq:CCG_A}
\end{eqnarray}
where $\va\vb$ denotes the dyadic product. The first two terms in Eq.~\eqref{eq:CCG_2ndMoments} respectively result from the weak measurement and classical fluctuations in the momentum displacements, while the anticommutator terms are exactly what one expects from a unitary pairwise interaction. Inserting the unregularized $\Phi_{nk}$ here would result in diverging diffusion rates $\Mat{A}_{nn}$ and is therefore not an option.
To resolve this problem, we propose to work with the Gaussian-regularized Newton potentials from the start, 
\begin{equation} \label{eq:Phink}
\Phi_{nk} (r) = -\frac{G m_n m_k}{r} \, \erf \left( \frac{r}{\sqrt{2} \sigma_{nk}} \right),
\end{equation}
as sourced from a Gaussian mass distribution with standard deviation $\sigma_{nk}$ and noting that a regularization is already present in the systematic dynamics. The mean force then takes on a very similar form with the same features,
\begin{equation} \label{eq:CCGforceReg}
\vF_n (\all{\vr}) =  \nabla_{\vr_n} \sum_{k\neq n} \frac{G m_n m_k}{r_{nk}} \erf \left( \frac{\vr_{nk}}{2 \sigma_{nk}} \right). 
\end{equation}
In the far-field limit, $r_{nk} \gg \sigma_{nk}$, it reduces to the Newton force, whereas in the near-field limit, the force becomes linear. Deviations from the Newton form are Gaussian suppressed in $r_{nk}/\sigma_{nk}$, as $\erf (x) \approx 1 - e^{-x^2}/\sqrt{\pi} x$, and thus experimentally accessible only at sufficiently large values of the CCG parameter $\sigma_0$.

The diffusion matrix $\Mat{A}_{nk}$, on the other hand, does not assume a simple form in general, but in the far-field limit it reduces to
\begin{equation} \label{eq:CCG_AfarField}
\Mat{A}_{nk} (\all{\vr}) \approx \frac{\vq_n (\all{\vr}) \vq_k (\all{\vr})}{\hbar^2} \approx \frac{G^2 m_n m_k}{\hbar^2 \gamma^2} \sum_{\substack{i\neq n \\ j\neq k}} m_i m_j \frac{\vr_{ni} \vr_{kj}}{r_{ni}^3 r_{kj}^3}. 
\end{equation}
This form is valid as long as no two masses are in close proximity (relative to $\sigma_{nk}$) to each other, and is therefore not applicable for composite particles.
Equations \eqref{eq:CCG_ME}, \eqref{eq:CCG_U} and \eqref{eq:Phink} define CCG for generic mechanical systems, generalizing the KTM model to arbitrary many-body configurations in the Newtonian regime.

The explicit consistency with KTM becomes evident in the condensed-matter scenario where the motion of all masses is bound to small deviations $\all{\ovr}$ from their respective equilibrium positions $\all{\vx}$. A second order expansion of \eqref{eq:CCG_ME} then leads to the diffusive master equation
\begin{eqnarray} \label{eq:CCG_ME2nd}
\cL_N \rho &\approx& \frac{i}{\hbar} \sum_n \left[ \left( 1 + \frac{1}{2} \sum_k \ovr_k \cdot \nabla_{\vx_k} \right) \vF_n (\all{\vx}) \cdot \ovr_n, \rho \right] \\
&&+ \sum_n \frac{\gamma}{4\sigma_n^2} \left[ \ovr_n \rho \cdot\ovr_n - \frac{\{ |\ovr_n|^2,\rho  \}}{2} \right] \nonumber \\
&&+ \gamma \sum_{n,k}  \left[ \ovr_n \rho \cdot \Mat{A}_{nk} (\all{\vx}) \ovr_k - \frac{\left\{ \ovr_n \cdot \Mat{A}_{nk} (\all{\vx}) \ovr_k, \rho \right\}}{2}  \right], \nonumber 
\end{eqnarray}
(see Appendix \ref{app:2ndorderME}). 
The first line in \eqref{eq:CCG_ME2nd} describes the effective gravitational interaction of each particle with the mean field of the others, whereas the second and third lines separate the CCG-induced diffusion due to position measurement and due to the applied feedback, respectively.
If we neglect the second order term in the first line, use the unregularized far-field Newton force, and insert the far-field approximation \eqref{eq:CCG_AfarField}, the equation recovers the KTM model for two or more masses at macroscopic distances \cite{Kafri2014,Khosla2017,Altamirano2018}.

\section{Observable consequences}\label{sec:observable}

We now survey the observable predictions of the CCG model, starting from the instructive example of two test masses $m_1 , m_2$. Specifically, let us focus on the impact on spatial coherence between two relative distance vectors $\vr,\vr'$ at a fixed center-of-mass position $\vR$. Assuming the particles are not too close, $r,r' \gg \sigma_{12}$, the corresponding matrix element evolves under CCG as
\begin{eqnarray}
&& \la \vr | \cL_2 \rho |\vr' \ra \approx - \gamma \la \vr | \rho |\vr' \ra  \label{eq:decoh_2masses}\\
&& \quad \times \left[ 1 - e^{-(\vr-\vr')^2/8\sigma_{12}^2 - 4i G m_1 m_2 (r^2-r'^2)/\gamma \hbar |\vr + \vr'|^3} \right]. \nonumber
\end{eqnarray}
This describes the simultaneous build-up of a coherent gravitational phase and dephasing at the rate $\gamma$. While the coherent phase ensures that the correct average Newton force emerges between the test masses, it is outpaced by the dephasing for superpositions over lengths $|\vr-\vr'| > \sigma_{12}$. This indicates that CCG-mediated gravity generally cannot entangle the motion of two test masses, consistent with the LOCC construction of the model.

\emph{Particle in earth's gravity.---}
For the motion of quantum particles in the vicinity of a macroscopic source of gravity, CCG causes not only decoherence, but also a small change in the background potential. 
Modern atom interferometers with large arm separations and long interrogation times should thus provide good bounds on the CCG parameters $\gamma,\sigma_0$, as they are able to detect smallest phase shifts due to fluctuations in the gravitational potential of the earth \cite{Peters2001,Rosi2014,Biedermann2015,Hamilton2015,Kovachy2015a,Jaffe2017}. At the same time, potential sources of dephasing can be excluded by maintaining a stable high interference contrast. Alternatively, spectroscopic experiments probing gravity with ultra-cold neutrons are a promising test bed \cite{Nesvizhevsky2002,Jenke2011,Abele2012,Cronenberg2015}.

To this end, consider a small test mass $m$ in the vicinity of a huge homogeneous sphere of mass $M\ggg m$ comprised of, say, identical mass elements $\delta m$. The sphere radius $R \ggg \sigma_s$ shall exceed by far the relevant standard deviation $\sigma_s = \sqrt{m_0 (m + \delta m)/m\delta m}\sigma_0$, while the position $\vr = z\ve_z$ of the test particle can be close to the surface, $z \gtrsim R$. By virtue of a continuum approximation, the average force \eqref{eq:CCGforceReg} on the particle becomes
\begin{equation}
F (z) \approx \partial_z \frac{G M m}{z} \left[ 1 + f(z) \right],
\end{equation}
to leading order in $\sigma_s/R$ (see App.~\ref{app:2spheres}). The deviation from the Newtonian form is of second order and described by the non-negative function
\begin{equation}
f(z) = \frac{z^3 + 3 R^2 z - 2 R^3}{4 R^3}\mbox{erfc} \left(\frac{z - R}{2 \sigma_s}\right) \leq 0.5,
\end{equation}
which vanishes exponentially as $z-R > \sigma_s$. Hence even for macroscopic hypothetical $\sigma_0$-values, the overall CCG-induced modification of the earth's homogeneous gravitational field may have no detectable impact on  trajectories as long as $\sigma_s \lll R$. The concomitant loss of interference contrast due to CCG-induced dephasing is capped at a maximum rate $\gamma$, regardless of the test mass. It is only in the diffusion limit, when the interference arm separation is smaller than the mass-rescaled standard deviation $\sigma_s$, that the reduced decoherence rate will depend on the particle mass, as described by the diffusive master equation \eqref{eq:CCG_ME2nd}.

\emph{Superposition of microspheres.---}
Finally, we address the recent proposal to test whether gravity is classical or quantum in a levitated optomechanics experiment with two microspheres \cite{Bose2017}. Suppose both spheres are prepared, in parallel, in a spatial superposition state where each arm corresponds to a different distance between them. If gravity were a quantum channel, the Newtonian attraction between the spheres should lead to the accumulation of a relative phase that entangles the two initially uncorrelated superposition states. 
Conversely, if gravity were classical, such a buildup of quantum correlation could not be observed. Hence the successful observation of some degree of entanglement between the spheres would put the CCG model to the test. 

For quantitative predictions, we consider the relative state of motion $\rho_r$ of two homogeneous rigid spheres $A$ and $B$ in the joint center-of-mass frame, with equal mass $M = N\delta m$ and radius $R$. 
For the spatial coherence between the relative coordinates $\vr$ or $\vr'$, CCG yields the rate of change
\begin{equation}
    \frac{\la \vr | \cL_{2N} \rho_r | \vr'\ra }{\la \vr | \rho_r | \vr'\ra } = \gamma \exp\left[-\tfrac{N |\vr-\vr'|^2}{16\sigma^2} \right] \mathcal{I} \left(\tfrac{\vr+\vr'}{2},\vr-\vr'\right) -\gamma, \label{eq:decoh_2spheres}
\end{equation}
with $\sigma = \sqrt{m_0/\delta m} \sigma_0$. 
Here the real part gives the actual rate of coherence decay, while the imaginary part describes the gravitational phase accumulation. Both are determined by a $6N$-dimensional Gaussian convolution integral of the regularized force kicks,
\begin{eqnarray}
\mathcal{I} (\vr_+,\vr_-) = \int \frac{\diff^{3N} \all{z}_A \diff^{3N} \all{z}_B}{(2\pi \sigma^2)^{3N}} e^{-\sum_{n=1}^N (\vz_{An}^2+\vz_{Bn}^2)/2\sigma^2} \label{eq:Idecoh_2spheres} \\
\times e^{- i\vr_- \cdot \nabla \sum_{n,k=1}^N \Phi_{nk} \left( \left| \vr_+ + \vx_{An} - \vx_{Bk} - \vz_{An} + \vz_{Bk}  \right| \right)/\hbar\gamma  } , \nonumber 
\end{eqnarray}
where $\all{\vx}_{A,B}$ denote the positions of the $N$ constituent masses $\delta m$ of each sphere relative to its center. 
The expression simplifies greatly in the regime where the CCG resolution is much smaller than the distance of the spheres, $\sigma \ll r_+$. The regularization of the Newton potential can then be neglected, and \eqref{eq:decoh_2spheres} reduces to the result \eqref{eq:decoh_2masses} for two point masses, with $m_{1,2}=M$ and $\sigma_{12} = \sqrt{2/N}\sigma$.
%\added[id=s]{[OPT OUT]}
%\begin{equation}
%\mathcal{I} (\vr_+,\vr_-) \xrightarrow{\sigma \ll R,r_+}    e^{-iGM^2 \vr_- \cdot \vr_+/\hbar \gamma r_+^3}.
%\end{equation}
Macroscopically distinct superpositions simply decohere at the full CCG rate $\gamma$, which fits into the picture that classical gravity does not mediate coherent phases. However, once the CCG resolution exceeds the arm separation, $\sigma > \sqrt{N} r_-$, a coherent phase could build up on top of the decoherence that occurs at a fraction of the rate $\gamma$. Note that this does not violate the LOCC-entanglement no-go theorem as the regime $\sigma > \sqrt{N}r_-$ implies the measurements are no longer local.

In fact, a key implication of CCG in the limit of poor resolution, $\sigma \gg r_+$, is a strongly suppressed gravity due to the regularization. Consider the first order expansion of \eqref{eq:Idecoh_2spheres} in the gravitational phase for homogeneous spheres (see App.~\ref{app:2spheres}),
\begin{eqnarray}
\mathcal{I} (\vr_+,\vr_-) &\approx& 1 + \frac{i \vr_- \cdot \vr_+}{\hbar \gamma r_+} \partial_{r_+} \left[ \frac{GM^2}{r_+} \mathcal{R} \left(\frac{\sigma}{R},\frac{r_+}{R} \right) \right], \label{eq:Idecoh_2spheres_approx} \\
\mathcal{R} (\alpha,\beta) &=& \frac{18}{\pi} \int_0^\infty \!\! \diff \xi  \frac{(\sin \xi - \xi \cos \xi)^2 \sin \beta \xi}{\xi^7} e^{-2\alpha^2 \xi^2}, \nonumber 
\end{eqnarray}
Here the square-bracketed term is the regularized potential between the two spheres at distance $r_+ > 2R$, with the Newtonian limit $\mathcal{R} (\alpha \ll \beta,\beta>2) \to 1$. In the opposite limit $\alpha \gg \beta$, the function $\mathcal{R}$ assumes much smaller values, and we may expand further to obtain
\begin{equation}
\mathcal{I} (\vr_+,\vr_-) \xrightarrow{\sigma \gg r_+} 1 - \frac{iG M^2 \vr_- \cdot \vr_+}{12\sqrt{2\pi}\hbar\gamma \sigma^3},
\end{equation}
a suppression of the Newtonian phase to third order in $r_+/\sigma$. 
This suggests that an exhaustive experimental test of classical gravity should not only focus on the gravitational decoherence effect, but also aim at detecting (or ruling out) systematic deviations from the Newtonian interaction potential.

\section{Conclusion}\label{sec:conclusion}

We have presented a consistent minimal model for the quantum evolution of matter under the influence of classical gravity in the Newtonian limit. It is formulated in terms of a Galilei-covariant measurement-feedback channel that applies universally to non-relativistic quantum many-body systems of all sizes. Avoiding the reference to a notoriously elusive quantized gravity, the model gives rise to an incoherent gravitational interaction between quantum test masses, under which the correct Newtonian force emerges at macroscopic distances.

Contrary to spontaneous collapse theories, whose main purpose is to establish macroscopic realism, the present model's decoherence effect does not amplify with the system size and thus leaves more room for quantum coherence on macroscopic scales. On the other hand, the model predicts a short-distance reqularization and dephasing of the gravitational interaction between macroscopic test masses that may soon be accessible in levitated optomechanics experiments.

\begin{acknowledgments}
KK and SN contributed equally to the concept and analysis, with the manuscript prepared by SN.
This research is supported by the Singapore Ministry of Education through the Academic Research Fund Tier 3 (Grant No. MOE2012-T3-1-009); and by the same MoE and the National Research Foundation, Prime Minister's Office, Singapore, under the Research Centres of Excellence programme, and by Australian Research Council Centre of Excellence for Engineered Quantum Systems (Grant No. CE110001013).
\end{acknowledgments}

\clearpage

\appendix

\begin{widetext}

\section{Master equation for the center of mass} \label{app:COM}

Here we show that the CCG master equation \eqref{eq:CCG_ME} reduces to the single-particle form \eqref{eq:CCG_ME1} for the center of mass of an isolated $N$-particle system. The $N$-particle master equation can be written as
\begin{eqnarray}
    \cL_N \rho &=& \gamma \prod_n \frac{m_n}{m_0} \int \frac{\diff^{3N} \all{z}}{(2\pi\sigma_0^2)^{3N/2}} \exp \left\{ - \sum_n \left[ \frac{m_n (\ovr_n - \vz_n)^2}{4m_0 \sigma_0^2} + \sum_{k\neq n} \frac{i \ovr_n \cdot (\vz_n - \vz_k) \Phi'_{nk} (|\vz_n-\vz_k|)}{\hbar \gamma |\vz_n - \vz_k|} \right] \right\} \rho \nonumber \\
    &&\times \exp \left\{ - \sum_n \left[ \frac{m_n (\ovr_n - \vz_n)^2}{4m_0 \sigma_0^2} - \sum_{k\neq n} \frac{i \ovr_n \cdot (\vz_n - \vz_k) \Phi'_{nk} (|\vz_n-\vz_k|)}{\hbar \gamma |\vz_n - \vz_k|} \right] \right\} - \gamma \rho.
\end{eqnarray}
In the center-of-mass frame, $\ovr_n = \ovR + \ovx_n$ with $\ovR = \sum_n m_n \ovr_n/M$ and $M=\sum_n m_n$. The $\ovx_n$ are linear combinations of the $N-1$ position operators of the relative coordinates, which turn into $\mathbb{C}$-vectors after tracing over them. In addition, $\Phi_{nk}(z) = \Phi_{kn}(z)$ and so the regularized force summed over all particles cancels as expected, $\sum_{n, k\neq n} \Phi_{nk}'(|\vz_n - \vz_k|) (\vz_n - \vz_k)/|\vz_n - \vz_k| = 0$. 
Hence the phase terms all cancel for the reduced center-of-mass state $\rho_{\rm cm} = \tr_{\rm rel} \rho$, and after substituting $\vz_n \to \vz_n - \vx_n$ we get
\begin{equation}
    \tr_{\rm rel} \cL_N \rho = \gamma \prod_{n=1}^N \frac{m_n}{m_0} \int \frac{\diff^{3N} \all{z}}{(2\pi\sigma_0^2)^{3N/2}} \exp \left[ - \sum_n \frac{ m_n (\ovR - \vz_n)^2}{4m_0 \sigma_0^2} \right] \rho_{\rm cm} \exp \left[ - \sum_n \frac{ m_n (\ovR - \vz_n)^2}{4m_0 \sigma_0^2} \right] - \gamma \rho_{\rm cm}. \label{eq:com_ME1}
\end{equation}
At this point, we must switch to relative and center-of-mass coordinates for the integration variables. For two coordinates and masses, we have the transformation rules and identities,
\begin{eqnarray}
    (m_a, \vz_a; m_b, \vz_b) &\to& \left( M_{ab}=m_a+m_b, \vZ_{ab} = \frac{m_a \vz_a + m_b \vz_b}{M_{ab}}; \mu_{a-b} = \frac{m_a m_b}{M_{ab}}, \vz_{a-b} = \vz_a - \vz_b \right), \nonumber \\
    &&\quad m_a z_a^2 + m_b z_b^2 = M_{ab}Z_{ab}^2 + \mu_{a-b}z_{a-b}^2, \quad \diff^3 z_a \diff^3 z_b = \diff^3 Z_{ab} \diff^3 z_{a-b}. \label{eq:comrelCoordRules} %\vz_a = \vz_{ab} + \frac{m_b}{M_{ab}}\vz_{a-b}, \quad \vz_b = \vz_{ab} - \frac{m_a}{M_{ab}}\vz_{a-b}
\end{eqnarray}
Iterating this over the $N$ coordinates $\all{\vz}$ results in the identities
\begin{equation}
    \sum_{n=1}^N m_n z_n^2 = M Z^2 + \sum_{j=2}^N \mu_{-j} z_{-j}^2, \quad \vZ = \frac{\sum_n m_n \vz_n}{M}, \quad \mu_{-j} = \frac{(m_1+\ldots + m_{j-1})m_j}{m_1 + \ldots + m_j}, \quad \diff^{3N} \all{z} = \diff^3 Z \prod_{j=2}^N \diff^3 z_{-j},
\end{equation}
which one can easily prove by induction. The Gaussian integrals over the $N-1$ relative coordinates $\vz_{-j}$ in \eqref{eq:com_ME1} can now be carried out explicitly,
\begin{equation}
    \tr_{\rm rel} \cL_N \rho = \frac{\gamma \prod_{n=1}^N m_n}{m_0 \prod_{j=2}^N \mu_{-j}} \int \frac{\diff^{3} Z}{(2\pi\sigma_0^2)^{3/2}} \exp \left[ - \frac{ M (\ovR - \vZ)^2}{4m_0 \sigma_0^2} \right] \rho_{\rm cm} \exp \left[ - \frac{ M (\ovR - \vZ)^2}{4m_0 \sigma_0^2} \right] - \gamma \rho_{\rm cm}. 
\end{equation}
Noticing further that $\prod_{j=2}^N \mu_{-j} = (\prod_{n=1}^N m_n)/M$ and expressing the Gaussian functions in terms of their Fourier transforms, we finally arrive at the single-particle form \eqref{eq:CCG_ME1} for $m=M$.

\section{Time evolution of first and second moments} \label{app:forceDiff}

For the equations of motion of the first and second moments, recall that the CCG Lindblad operators $\oL (\all{\vz}) = \oU(\all{\vz}) \mathcal{M} (\all{\ovr} - \all{\vz})$ are diagonal in position representation, and $\int\diff^{3N}\all{z} \oL^\da (\all{\vz}) \oL (\all{\vz}) = 1$. Using also that $[\ovp, f(\ovr)] = -i\hbar\nabla f(\ovr)$, the average momentum of a particle changes due to CCG as
\begin{eqnarray}
    \partial_{t,\rm CCG} \qerw{\ovp_n} &\equiv& \qerw{\vF_n (\all{\ovr})} = \qerw{\cL_N^\da \ovp_n} = \gamma \qerw{\int\diff^{3N} \all{z} \oL^\da (\all{\vz}) \left[ \ovp_n, \oL (\all{\vz}) \right]} =  -i\hbar \gamma \qerw{\int\diff^{3N} \all{z} \oL^\da (\all{\vz}) \nabla_{\ovr_n} \oL (\all{\vz}) } \nonumber \\
    &=& i\hbar \gamma \qerw{\int\diff^{3N} \all{z} \mathcal{M}^2 (\all{\ovr}-\all{\vz}) \left[ \frac{m_n (\ovr_n - \vz_n)}{2m_0\sigma_0^2} - \frac{i}{\hbar} \vq_n (\all{\vz}) \right]} = - \sum_{k\neq n} \qerw{\int\diff^{3N} \all{z} \mathcal{M}^2 (\all{\vz}) \nabla \Phi_{nk} (|\ovr_{nk} - \vz_n + \vz_k|)} \nonumber \\
    &=& - \sum_{k\neq n} \qerw{\nabla_{\ovr_n} \int \frac{(m_n m_k)^{3/2}\diff^3 z_n \diff^3 z_k}{(2\pi m_0 \sigma_0^2)^{3}} e^{-(m_n z_n^2 + m_k z_k^2)/2m_0 \sigma_0^2} \Phi_{nk} (|\ovr_{nk} - \vz_n + \vz_k|)}. \label{eq:app_force}
\end{eqnarray}
In the third line, we have carried out all $N-2$ trivial Gaussian integrals. Another such simplification can be done by switching to center-of-mass and relative coordinates for the remaining two variables $\vz_n,\vz_k$ using \eqref{eq:comrelCoordRules}, after which we recover the expression \eqref{eq:CCGforce1} in the main text.

The calculation for the second moments is slightly more tedious, but follows similar steps as above. For a convenient notation, we label the Cartesian components with Greek indices,
\begin{eqnarray}
    \partial_{t,\rm CCG} \qerw{\op_{n\alpha} \op_{k\beta}} &=& \gamma \qerw{\int\diff^{3N} \all{z} \oL^\da (\all{\vz}) \left[\op_{n\alpha} \op_{k\beta}, \oL (\all{\vz}) \right]} =  -i\hbar \gamma \qerw{\int\diff^{3N} \all{z} \oL^\da (\all{\vz}) \left[ \op_{n\alpha} \frac{\partial \oL (\all{\vz})}{\partial \ox_{k\beta}} + \frac{\partial \oL (\all{\vz})}{\partial \ox_{n\alpha}} \op_{k\beta} \right] } \nonumber \\
    &=&  -i\hbar \gamma \int\diff^{3N} \all{z} \qerw{ \op_{n\alpha} \oL^\da (\all{\vz})  \frac{\partial \oL (\all{\vz})}{\partial \ox_{k\beta}} + \oL^\da (\all{\vz}) \frac{\partial \oL (\all{\vz})}{\partial \ox_{n\alpha}} \op_{k\beta} + i\hbar \frac{\partial \oL^\da (\all{\vz})}{\partial \ox_{n\alpha}} \frac{\partial \oL (\all{\vz})}{\partial \ox_{k\beta}} } \nonumber \\
    &=& \gamma \int\diff^{3N} \all{z} \left\la \op_{n\alpha} \mathcal{M}^2 (\all{\ovr}-\all{\vz}) \left[i\hbar \frac{\ox_{k\beta}-z_{k\beta}}{2\sigma_k^2} + q_{k\beta} (\all{\vz}) \right] +  \mathcal{M}^2 (\all{\ovr}-\all{\vz}) \left[i\hbar \frac{\ox_{n\alpha}-z_{n\alpha}}{2\sigma_n^2} + q_{n\alpha} (\all{\vz}) \right] \op_{k\beta} \right. \nonumber \\
    && \left. + \mathcal{M}^2 (\all{\ovr}-\all{\vz}) \left[i\hbar \frac{\ox_{n\alpha}-z_{n\alpha}}{2\sigma_n^2} + q_{n\alpha} (\all{\vz}) \right]^\da \left[i\hbar \frac{\ox_{k\beta}-z_{k\beta}}{2\sigma_k^2} + q_{k\beta} (\all{\vz}) \right]  \right\ra . 
\end{eqnarray}
Now we can exploit once again that the Gaussian integrals over first- and mixed second-order terms in the components $\ox_{n\alpha} - z_{n\alpha}$ vanish, whereas the integral over $(\ox_{n\alpha} - z_{n\alpha})^2$ yields $\sigma_n^2$. Moreover, $\vF_n (\all{\ovr}) = \gamma \int\diff^{3N} \all{z} \mathcal{M}^2 (\all{\ovr}-\all{\vz}) \vq_n (\all{\vz})$ in \eqref{eq:app_force}, and so
\begin{eqnarray}
     \partial_{t,\rm CCG} \qerw{\op_{n\alpha} \op_{k\beta}} &=& \qerw{\op_{n\alpha} F_{k\beta} (\all{\ovr}) + F_{n\alpha} (\all{\ovr}) \op_{k\beta}} + \frac{\hbar^2 \gamma}{4\sigma_n^2} \delta_{nk}\delta_{\alpha \beta} + \gamma \int\diff^{3N} \all{z} \qerw{ \mathcal{M}^2 (\all{\ovr}-\all{\vz})  q_{n\alpha} (\all{\vz})  q_{k\beta} (\all{\vz}) } \nonumber \\
     && + i\hbar \gamma \int\diff^{3N} \all{z} \qerw{ \mathcal{M}^2 (\all{\ovr}-\all{\vz}) \left[ \frac{\ox_{k\beta} - z_{k\beta}}{2\sigma_k^2} q_{n\alpha} (\all{\vz}) - \frac{\ox_{n\alpha} - z_{n\alpha}}{2\sigma_n^2} q_{k\beta} (\all{\vz}) \right] }.
\end{eqnarray}
In the first line, we identify the last term as a matrix element of $\Mat{A}_{nk}$, see \eqref{eq:CCG_A} in the main text. For the second line, we make use of the identity $\partial \mathcal{M}^2 (\all{\vz})/\partial z_{n\alpha} = - z_{n\alpha}\mathcal{M}^2/\sigma_n^2$, so that
\begin{eqnarray}
     \partial_{t,\rm CCG} \qerw{\op_{n\alpha} \op_{k\beta}} &=& \qerw{\op_{n\alpha} F_{k\beta} (\all{\ovr}) + F_{n\alpha} (\all{\ovr}) \op_{k\beta}} + \frac{\hbar^2 \gamma}{4\sigma_n^2} \delta_{nk}\delta_{\alpha \beta} + \gamma \hbar^2 \qerw{ \left[ \Mat{A}_{nk} (\all{\ovr}) \right]_{\alpha \beta} }  + \frac{i\hbar}{2} \qerw{\frac{\partial F_{k\beta} (\all{\ovr})}{\partial \ox_{n\alpha}} - \frac{\partial F_{n\alpha} (\all{\ovr})}{\partial \ox_{k\beta}} } \nonumber \\
     &=& \frac{1}{2}\qerw{  \left\{ \op_{n\alpha}, F_{k\beta} (\all{\ovr}) \right\} + \left\{ \op_{k\beta}, F_{n\alpha} (\all{\ovr}) \right\} }  + \frac{\hbar^2 \gamma}{4\sigma_n^2} \delta_{nk}\delta_{\alpha \beta} + \gamma \hbar^2 \qerw{ \left[ \Mat{A}_{nk} (\all{\ovr}) \right]_{\alpha \beta} }.
\end{eqnarray}
Written in dyadic matrix notation, this reduces to \eqref{eq:CCG_2ndMoments} in the main text.

\section{Second order expansion of the master equation} \label{app:2ndorderME}

In many-body configurations where the motion of individual test masses is spatially confined and where the typical distance to neighbouring test masses greatly exceeds that confinement, the CCG master equation reduces to the linearized KTM model. To see this explicitly, we shall redefine the position operator of each particle relative to its equilibrium position, $\ovr_n \to \vx_n + \ovr_n$, and expand the master equation \eqref{eq:CCG_ME} to second order in deviations from the $\vx_n$. Notations and identities of App.~\ref{app:forceDiff} will be reused. The CCG Lindblad operators expand as
\begin{eqnarray}
     \oL (\all{\vz}) &\approx&  \mathcal{M}(\all{\vx}-\all{\vz}) \prod_n \left\{ 1 + \left( \frac{i\vq_n (\all{\vz})}{\hbar} + \frac{\vz_n - \vx_n}{2\sigma_n^2} \right)\cdot \ovr_n + \frac{1}{2} \left[ \left( \frac{i\vq_n (\all{\vz})}{\hbar} + \frac{\vz_n - \vx_n}{2\sigma_n^2} \right)\cdot \ovr_n \right]^2 - \frac{\ovr_n^2}{4\sigma_n^2} \right\} \nonumber \\
     &\approx& \mathcal{M}(\all{\vx}-\all{\vz}) \left\{ 1 + \sum_n \left[ \left( \frac{i\vq_n (\all{\vz})}{\hbar} + \frac{\vz_n - \vx_n}{2\sigma_n^2} \right)\cdot \ovr_n - \frac{\ovr_n^2}{4\sigma_n^2} \right] + \frac{1}{2} \left[ \sum_n \left( \frac{i\vq_n (\all{\vz})}{\hbar} + \frac{\vz_n - \vx_n}{2\sigma_n^2} \right)\cdot \ovr_n \right]^2  \right\}
\end{eqnarray}
where the phase factor has been dropped. To second order, these operators remain a partition of the identity, 
\begin{eqnarray}
     \oL^\da \oL (\all{\vz}) &\approx& \mathcal{M}^2 (\all{\vx}-\all{\vz}) \left\{ 1 + \sum_n \frac{2(\vz_n - \vx_n)\cdot\ovr_n - \ovr_n^2 }{2\sigma_n^2} + 2 \left[ \sum_n \frac{(\vz_n - \vx_n)\cdot\ovr_n}{2\sigma_n^2} \right]^2  \right\},\quad 
     \int\diff^{3N} \all{z} \, \oL^\da \oL (\all{\vz}) \approx \id .
\end{eqnarray}
Hence we can insert the expansion into the CCG master equation. Carrying out some of the trivial Gaussian integrals, we obtain to second order, 
\begin{eqnarray}
     \cL_N \rho &\approx& \gamma \int \diff^{3N} \all{z} \, \mathcal{M}^2 (\all{\vx}-\all{\vz}) \left\{ \sum_n \left( \frac{i}{\hbar} [\vq_n (\all{\vz})\cdot\ovr_n,\rho ] - \frac{\{ \ovr_n^2, \rho\}}{4\sigma_n^2} \right) + \frac{1}{2} \left( \left[ \sum_n \left( \frac{i\vq_n (\all{\vz})}{\hbar} + \frac{\vz_n - \vx_n}{2\sigma_n^2} \right)\cdot \ovr_n \right]^2 \rho + h.c. \right)  \right. \nonumber \\
     && \left. + \left[ \sum_n \left( \frac{i\vq_n (\all{\vz})}{\hbar} + \frac{\vz_n - \vx_n}{2\sigma_n^2} \right)\cdot \ovr_n \right] \rho \left[ \sum_k \left( -\frac{i\vq_k (\all{\vz})}{\hbar} + \frac{\vz_k - \vx_k}{2\sigma_k^2} \right)\cdot \ovr_k \right] \right\} \nonumber \\
     &=& \frac{i}{\hbar} \left[ \sum_{n,\alpha} F_{n\alpha} (\all{\vx}) \ox_{n\alpha}, \rho \right] - \sum_{n,\alpha} \frac{\gamma}{8\sigma_n^2} [\ox_{n\alpha}, [\ox_{n\alpha}, \rho]] + \gamma \sum_{\substack{n,\alpha \\ k,\beta}} \left[ \Mat{A}_{nk} (\all{\vx})  \right]_{\alpha \beta} \left[ \ox_{n\alpha} \rho \ox_{k\beta} - \frac{\left\{ \ox_{n\alpha} \ox_{k\beta}, \rho \right\}}{2} \right] \nonumber \\
     &&+\frac{i\gamma}{\hbar} \sum_{\substack{n,\alpha \\ k,\beta}} \int \diff^{3N} \all{z} \, \mathcal{M}^2 (\all{\vx}-\all{\vz}) \left\{ \left[ q_{n\alpha} (\all{\vz}) \frac{z_{k\beta}-x_{k\beta}}{2\sigma_k^2} - q_{k\beta} (\all{\vz}) \frac{z_{n\alpha}-x_{n\alpha}}{2\sigma_n^2} \right] \ox_{n\alpha} \rho \ox_{k\beta} + q_{n\alpha} (\all{\vz}) \frac{z_{k\beta}-x_{k\beta}}{2\sigma_k^2} \left[ \ox_{n\alpha} \ox_{k\beta},\rho \right] \right\} \nonumber \\
     &=& \frac{i}{\hbar} \left[ \sum_{n,\alpha} F_{n\alpha} (\all{\vx}) \ox_{n\alpha}, \rho \right] - \sum_{n,\alpha} \frac{\gamma}{8\sigma_n^2} [\ox_{n\alpha}, [\ox_{n\alpha}, \rho]] + \gamma \sum_{\substack{n,\alpha \\ k,\beta}} \left[ \Mat{A}_{nk} (\all{\vx})  \right]_{\alpha \beta} \left[ \ox_{n\alpha} \rho \ox_{k\beta} - \frac{\left\{ \ox_{n\alpha} \ox_{k\beta}, \rho \right\}}{2} \right] \nonumber \\
     &&+\frac{i}{2\hbar} \sum_{\substack{n,\alpha \\ k,\beta}} \left[ \frac{\partial F_{n\alpha} (\all{\vx})}{\partial x_{k\beta}} \ox_{n\alpha} \ox_{k\beta}, \rho \right] + \frac{i}{2\hbar} \sum_{\substack{n,\alpha \\ k,\beta}} \left[ \frac{\partial F_{n\alpha} (\all{\vx})}{\partial x_{k\beta}} - \frac{\partial F_{k\beta} (\all{\vx})}{\partial x_{n\alpha}} \right] \ox_{n\alpha} \rho \ox_{k\beta} .
\end{eqnarray}
We are now one term away from the final result, \eqref{eq:CCG_ME2nd} in the main text. It remains to be shown that the last term in the last line of the above expression vanishes, i.e.~$\partial F_{n\alpha}/\partial x_{k\beta} = \partial F_{k\beta} / \partial x_{n\alpha} $. To this end, we employ the explicit form \eqref{eq:CCGforce1} for the force $\vF_n (\all{\vx})$ and note that we can include $n$ in the summation there without changing the result. As a result,
\begin{eqnarray}
     \frac{\partial F_{n\alpha} (\all{\vx})}{\partial x_{k\beta}} - \frac{\partial F_{k\beta} (\all{\vx})}{\partial x_{n\alpha}} &=& \frac{\partial^2}{\partial x_{n\alpha} \partial x_{k\beta}} \sum_{\ell} \int\diff^3 z \left[ \frac{e^{-z^2/2\sigma_{n\ell}^2}}{(2\pi\sigma_{n\ell}^2)^{3/2}} \Phi_{n\ell} (|\vx_n - \vx_\ell - \vz|) - \frac{e^{-z^2/2\sigma_{k\ell}^2}}{(2\pi\sigma_{k\ell}^2)^{3/2}} \Phi_{k\ell} (|\vx_k - \vx_\ell - \vz|) \right] \nonumber \\
     &=& \frac{\partial^2}{\partial x_{n\alpha} \partial x_{k\beta}} \int\diff^3 z \left[ \frac{e^{-z^2/2\sigma_{nk}^2}}{(2\pi\sigma_{nk}^2)^{3/2}} \Phi_{nk} (|\vx_n - \vx_k - \vz|) - \frac{e^{-z^2/2\sigma_{kn}^2}}{(2\pi\sigma_{kn}^2)^{3/2}} \Phi_{kn} (|\vx_k - \vx_n - \vz|) \right].
\end{eqnarray}
It follows immediately that this expression vanishes if $k=n$. For all other $k\neq n$, we arrive at the same conclusion after making use of the symmetries $\sigma_{nk}=\sigma_{kn}$ and $\Phi_{nk} (z) = \Phi_{kn}(z)$. This is intuitively understood by considering action-reaction pairs with motion in different Cartesian directions geometrically contributing to the $1/r$ potential.

\section{Effective force between two homogeneous spheres} \label{app:2spheres}

Here we show explicitly how to obtain the approximate phase modulation \eqref{eq:Idecoh_2spheres_approx} between two homogeneous spheres from the full expression \eqref{eq:Idecoh_2spheres}, followed by the effective potential of a test particle near to a massive object. The first step is a Taylor expansion of the phase factor which, by virtue of the definition \eqref{eq:CCGforce1} and the explicit form \eqref{eq:CCGforceReg} for the regularized forces, leaves us with $\mathcal{I} (\vr_+,\vr_-) \approx 1 + i \vr_- \cdot \nabla \Phi_{\rm eff} (\vr_+)/\hbar \gamma $ and 
\begin{equation}
    \Phi_{\rm eff} (\vr) = \sum_{n,k=1}^N \frac{G \delta m^2}{|\vr + \vx_{An}-\vx_{Bk}|} \erf \left( \frac{|\vr + \vx_{An}-\vx_{Bk}|}{2\sqrt{2}\sigma} \right) \approx G \left( \frac{3M}{4\pi R^3}\right)^2 \iint_{\substack{x_A \leq R \\ x_B \leq R}} \diff^3 x_A \diff^3 x_B \frac{\erf (|\vr + \vx_A - \vx_B|/2\sqrt{2} \sigma)}{|\vr + \vx_A - \vx_B|}.
\end{equation}
Here, we have made the continuum approximation for homogeneous spheres of radius $R$ and mass $M$. Given that the 3D-Fourier transform of $\erf(r/\sqrt{2}a)/r$ is $4\pi \exp(-a^2 k^2/2)/k^2$, we can further simplify
\begin{equation}
    \Phi_{\rm eff} (\vr) =  \frac{G}{2\pi^2} \left( \frac{3M}{4\pi R^3}\right)^2 \int \frac{\diff^3 k}{k^2} e^{-2\sigma^2 k^2 + i\vk\cdot\vr}  \left| \int_{x\leq R} \diff^3 x \, e^{i\vk \cdot \vx} \right|^2 = \frac{2G}{\pi} \left( \frac{3M}{4\pi R^3}\right)^2 \int_0^\infty \diff k \, e^{-2\sigma^2 k^2} \sinc (k r) h^2(kR).
\end{equation}
This expression leads to \eqref{eq:Idecoh_2spheres_approx} in the main text once we insert
\begin{equation}
    h(k R) = \int_{x\leq R} \diff^3 x \, e^{i\vk \cdot \vx} = \frac{4\pi}{k^3}\left( \sin kR - kR \cos kR \right).
\end{equation}

To compute the effective potential between a large spherical mass and test particle at distance $z \gtrsim R$ we again make the continuum approximation and integrate over the sphere. This can be done explicitly from the regularized potential in Eq.~\eqref{eq:CCGforceReg}. Setting now $\sigma_s = \sqrt{m_0 (m+\delta m)/m\delta m} \sigma_0$,  

\begin{eqnarray}
     \Phi_{\rm eff} (z) &=& \frac{GMm}{V} \int_{r<R} \diff^3 r \frac{\erf (|\vr - z\vec{e}_z|/2\sigma_s)}{|\vr - z\vec{e}_z|} = \frac{GMm}{V} \int_0^R r^2 \diff r \int_{-1}^1 \diff \xi \, 2\pi \frac{\erf (\sqrt{r^2+ z^2 - 2rz\xi}/2\sigma_s)}{\sqrt{r^2+ z^2 - 2rz\xi}} \\
     &=& \frac{2\pi G M m }{V} \int_0^R \diff r \, \frac{r}{z} \left[ - (z-r)\erf \left( \frac{z-r}{2\sigma_s} \right) - \frac{2\sigma_s}{\sqrt{\pi}} e^{-(r-z)^2/4\sigma_s^2} + (z+r)\erf \left( \frac{z+r}{2\sigma_s} \right) + \frac{2\sigma_s}{\sqrt{\pi}} e^{-(r+z)^2/4\sigma_s^2}  \right]. \nonumber
\end{eqnarray}
Here, we can drop the last Gaussian term in the integral and replace the second error function by unity, since $r+z \sim R \ggg \sigma_s$. The remaining terms can be integrated explicitly, 
\begin{eqnarray}
     \Phi_{\rm eff} (z) &=& \frac{2\pi G M m }{V z} \int_0^R \diff r \, \left[ 2r^2 - r(z-r) \left[ \erf \left( \frac{z-r}{2\sigma_s}\right) - 1 \right] - \frac{2r\sigma_s}{\sqrt{\pi}} e^{-(r-z)^2/4\sigma_s^2} \right] \\
     &=& \frac{GmM}{z} \left\{ 1 - \frac{3}{2R^3}\int_0^R \diff r \left[ r(z-r) \left[ \erf \left( \frac{z-r}{2\sigma_s}\right) - 1 \right] + \frac{2r\sigma_s}{\sqrt{\pi}} e^{-(r-z)^2/4\sigma_s^2} \right] \right\} \nonumber \\
     &=& \frac{GmM}{z} \left\{ 1 - \frac{3}{2R^3} \left[ \left( \frac{zR^2}{2} - \frac{R^3}{3} \right) \left[ \erf \left( \frac{z-R}{2\sigma_s}\right) - 1 \right] + \int_0^R \diff r \frac{r}{\sqrt{\pi}} e^{-(r-z)^2/4\sigma_s^2} \left( 2\sigma_s - \frac{3zr-2r^2}{6\sigma_s} \right) \right] \right\} . \nonumber
\end{eqnarray}
To get to the last line, partial integration was done on the error function term. The remaining Gaussian integral can be carried out explicitly. Keeping only terms linear in $\sigma_s/R$, and making the same Gaussian and error function approximations as above we find, 
\begin{eqnarray}
     \Phi_{\rm eff} (z) &=& \frac{GmM}{z} \left\{ 1 + \frac{z^3 + 3 R^2 z - 2 R^3}{4 R^3}\left[1 - \erf \left(\frac{z - R}{2 \sigma_s}\right) \right]  -  \frac{\sigma_s  (z-R) (2 R+z)}{2\sqrt{\pi}R^3}e^{-(z - R)^2/4\sigma_s^2} \right\}.
\end{eqnarray}
We note the Gaussian term is suppressed by $\sigma_s/R$ even for $z = R$, while the 1-erf term has a maximum of $\frac12$ at $z = R$, but is also Gaussian suppressed for $z - R \gtrsim \sigma_s$.

\end{widetext}

\bibliography{ccg}

%merlin.mbs apsrev4-1.bst 2010-07-25 4.21a (PWD, AO, DPC) hacked
%Control: key (0)
%Control: author (8) initials jnrlst
%Control: editor formatted (1) identically to author
%Control: production of article title (-1) disabled
%Control: page (0) single
%Control: year (1) truncated
%Control: production of eprint (0) enabled
\begin{thebibliography}{50}%
\makeatletter
\providecommand \@ifxundefined [1]{%
 \@ifx{#1\undefined}
}%
\providecommand \@ifnum [1]{%
 \ifnum #1\expandafter \@firstoftwo
 \else \expandafter \@secondoftwo
 \fi
}%
\providecommand \@ifx [1]{%
 \ifx #1\expandafter \@firstoftwo
 \else \expandafter \@secondoftwo
 \fi
}%
\providecommand \natexlab [1]{#1}%
\providecommand \enquote  [1]{``#1''}%
\providecommand \bibnamefont  [1]{#1}%
\providecommand \bibfnamefont [1]{#1}%
\providecommand \citenamefont [1]{#1}%
\providecommand \href@noop [0]{\@secondoftwo}%
\providecommand \href [0]{\begingroup \@sanitize@url \@href}%
\providecommand \@href[1]{\@@startlink{#1}\@@href}%
\providecommand \@@href[1]{\endgroup#1\@@endlink}%
\providecommand \@sanitize@url [0]{\catcode `\\12\catcode `\$12\catcode
  `\&12\catcode `\#12\catcode `\^12\catcode `\_12\catcode `\%12\relax}%
\providecommand \@@startlink[1]{}%
\providecommand \@@endlink[0]{}%
\providecommand \url  [0]{\begingroup\@sanitize@url \@url }%
\providecommand \@url [1]{\endgroup\@href {#1}{\urlprefix }}%
\providecommand \urlprefix  [0]{URL }%
\providecommand \Eprint [0]{\href }%
\providecommand \doibase [0]{http://dx.doi.org/}%
\providecommand \selectlanguage [0]{\@gobble}%
\providecommand \bibinfo  [0]{\@secondoftwo}%
\providecommand \bibfield  [0]{\@secondoftwo}%
\providecommand \translation [1]{[#1]}%
\providecommand \BibitemOpen [0]{}%
\providecommand \bibitemStop [0]{}%
\providecommand \bibitemNoStop [0]{.\EOS\space}%
\providecommand \EOS [0]{\spacefactor3000\relax}%
\providecommand \BibitemShut  [1]{\csname bibitem#1\endcsname}%
\let\auto@bib@innerbib\@empty
%</preamble>
\bibitem [{\citenamefont {Pikovski}\ \emph {et~al.}(2012)\citenamefont
  {Pikovski}, \citenamefont {Vanner}, \citenamefont {Aspelmeyer}, \citenamefont
  {Kim},\ and\ \citenamefont {Brukner}}]{Pikovski2012}%
  \BibitemOpen
  \bibfield  {author} {\bibinfo {author} {\bibfnamefont {I.}~\bibnamefont
  {Pikovski}}, \bibinfo {author} {\bibfnamefont {M.~R.}\ \bibnamefont
  {Vanner}}, \bibinfo {author} {\bibfnamefont {M.}~\bibnamefont {Aspelmeyer}},
  \bibinfo {author} {\bibfnamefont {M.~S.}\ \bibnamefont {Kim}}, \ and\
  \bibinfo {author} {\bibfnamefont {{\v{C}}.}~\bibnamefont {Brukner}},\ }\href
  {\doibase 10.1038/nphys2262} {\bibfield  {journal} {\bibinfo  {journal} {Nat.
  Phys.}\ }\textbf {\bibinfo {volume} {8}},\ \bibinfo {pages} {393} (\bibinfo
  {year} {2012})}\BibitemShut {NoStop}%
\bibitem [{\citenamefont {Pfister}\ \emph {et~al.}(2016)\citenamefont
  {Pfister}, \citenamefont {Kaniewski}, \citenamefont {Tomamichel},
  \citenamefont {Mantri}, \citenamefont {Schmucker}, \citenamefont {McMahon},
  \citenamefont {Milburn},\ and\ \citenamefont {Wehner}}]{Pfister2016}%
  \BibitemOpen
  \bibfield  {author} {\bibinfo {author} {\bibfnamefont {C.}~\bibnamefont
  {Pfister}}, \bibinfo {author} {\bibfnamefont {J.}~\bibnamefont {Kaniewski}},
  \bibinfo {author} {\bibfnamefont {M.}~\bibnamefont {Tomamichel}}, \bibinfo
  {author} {\bibfnamefont {A.}~\bibnamefont {Mantri}}, \bibinfo {author}
  {\bibfnamefont {R.}~\bibnamefont {Schmucker}}, \bibinfo {author}
  {\bibfnamefont {N.}~\bibnamefont {McMahon}}, \bibinfo {author} {\bibfnamefont
  {G.}~\bibnamefont {Milburn}}, \ and\ \bibinfo {author} {\bibfnamefont
  {S.}~\bibnamefont {Wehner}},\ }\href {\doibase 10.1038/ncomms13022}
  {\bibfield  {journal} {\bibinfo  {journal} {Nat. Commun.}\ }\textbf {\bibinfo
  {volume} {7}},\ \bibinfo {pages} {13022} (\bibinfo {year}
  {2016})}\BibitemShut {NoStop}%
\bibitem [{\citenamefont {Marletto}\ and\ \citenamefont
  {Vedral}(2017)}]{Marletto2017}%
  \BibitemOpen
  \bibfield  {author} {\bibinfo {author} {\bibfnamefont {C.}~\bibnamefont
  {Marletto}}\ and\ \bibinfo {author} {\bibfnamefont {V.}~\bibnamefont
  {Vedral}},\ }\href {\doibase 10.1103/PhysRevLett.119.240402} {\bibfield
  {journal} {\bibinfo  {journal} {Phys. Rev. Lett.}\ }\textbf {\bibinfo
  {volume} {119}},\ \bibinfo {pages} {240402} (\bibinfo {year}
  {2017})}\BibitemShut {NoStop}%
\bibitem [{\citenamefont {Bose}\ \emph {et~al.}(2017)\citenamefont {Bose},
  \citenamefont {Mazumdar}, \citenamefont {Morley}, \citenamefont {Ulbricht},
  \citenamefont {Toro{\v{s}}}, \citenamefont {Paternostro}, \citenamefont
  {Geraci}, \citenamefont {Barker}, \citenamefont {Kim},\ and\ \citenamefont
  {Milburn}}]{Bose2017}%
  \BibitemOpen
  \bibfield  {author} {\bibinfo {author} {\bibfnamefont {S.}~\bibnamefont
  {Bose}}, \bibinfo {author} {\bibfnamefont {A.}~\bibnamefont {Mazumdar}},
  \bibinfo {author} {\bibfnamefont {G.~W.}\ \bibnamefont {Morley}}, \bibinfo
  {author} {\bibfnamefont {H.}~\bibnamefont {Ulbricht}}, \bibinfo {author}
  {\bibfnamefont {M.}~\bibnamefont {Toro{\v{s}}}}, \bibinfo {author}
  {\bibfnamefont {M.}~\bibnamefont {Paternostro}}, \bibinfo {author}
  {\bibfnamefont {A.~A.}\ \bibnamefont {Geraci}}, \bibinfo {author}
  {\bibfnamefont {P.~F.}\ \bibnamefont {Barker}}, \bibinfo {author}
  {\bibfnamefont {M.~S.}\ \bibnamefont {Kim}}, \ and\ \bibinfo {author}
  {\bibfnamefont {G.}~\bibnamefont {Milburn}},\ }\href {\doibase
  10.1103/PhysRevLett.119.240401} {\bibfield  {journal} {\bibinfo  {journal}
  {Phys. Rev. Lett.}\ }\textbf {\bibinfo {volume} {119}},\ \bibinfo {pages}
  {240401} (\bibinfo {year} {2017})}\BibitemShut {NoStop}%
\bibitem [{\citenamefont {Carney}\ \emph {et~al.}(2018)\citenamefont {Carney},
  \citenamefont {Stamp},\ and\ \citenamefont {Taylor}}]{Carney2018}%
  \BibitemOpen
  \bibfield  {author} {\bibinfo {author} {\bibfnamefont {D.}~\bibnamefont
  {Carney}}, \bibinfo {author} {\bibfnamefont {P.~C.~E.}\ \bibnamefont
  {Stamp}}, \ and\ \bibinfo {author} {\bibfnamefont {J.~M.}\ \bibnamefont
  {Taylor}},\ }\href {http://arxiv.org/abs/1807.11494} {\  (\bibinfo {year}
  {2018})},\ \Eprint {http://arxiv.org/abs/1807.11494} {arXiv:1807.11494}
  \BibitemShut {NoStop}%
\bibitem [{\citenamefont {Rovelli}(2004)}]{Rovelli2004}%
  \BibitemOpen
  \bibfield  {author} {\bibinfo {author} {\bibfnamefont {C.}~\bibnamefont
  {Rovelli}},\ }\href {\doibase 10.1017/CBO9780511755804} {\emph {\bibinfo
  {title} {{Quantum gravity}}}}\ (\bibinfo  {publisher} {Cambridge University
  Press},\ \bibinfo {address} {Cambridge},\ \bibinfo {year} {2004})\BibitemShut
  {NoStop}%
\bibitem [{\citenamefont {Oriti}(2009)}]{Oriti2009}%
  \BibitemOpen
  \bibinfo {editor} {\bibfnamefont {D.}~\bibnamefont {Oriti}},\ ed.,\ \href
  {\doibase 10.1017/CBO9780511575549} {\emph {\bibinfo {title} {{Approaches to
  Quantum Gravity}}}}\ (\bibinfo  {publisher} {Cambridge University Press},\
  \bibinfo {address} {Cambridge},\ \bibinfo {year} {2009})\BibitemShut
  {NoStop}%
\bibitem [{\citenamefont {Kiefer}(2012)}]{Kiefer2012}%
  \BibitemOpen
  \bibfield  {author} {\bibinfo {author} {\bibfnamefont {C.}~\bibnamefont
  {Kiefer}},\ }\href {\doibase 10.1093/acprof:oso/9780199585205.001.0001}
  {\emph {\bibinfo {title} {{Quantum gravity}}}}\ (\bibinfo  {publisher}
  {Oxford University Press},\ \bibinfo {address} {Oxford},\ \bibinfo {year}
  {2012})\BibitemShut {NoStop}%
\bibitem [{\citenamefont {Karolyhazy}(1966)}]{Karolyhazy1966}%
  \BibitemOpen
  \bibfield  {author} {\bibinfo {author} {\bibfnamefont {F.}~\bibnamefont
  {Karolyhazy}},\ }\href {\doibase 10.1007/BF02717926} {\bibfield  {journal}
  {\bibinfo  {journal} {Nuovo Cim. A}\ }\textbf {\bibinfo {volume} {42}},\
  \bibinfo {pages} {390} (\bibinfo {year} {1966})}\BibitemShut {NoStop}%
\bibitem [{\citenamefont {Di{\'{o}}si}(1987)}]{Diosi1987}%
  \BibitemOpen
  \bibfield  {author} {\bibinfo {author} {\bibfnamefont {L.}~\bibnamefont
  {Di{\'{o}}si}},\ }\href {\doibase 10.1016/0375-9601(87)90681-5} {\bibfield
  {journal} {\bibinfo  {journal} {Phys. Lett. A}\ }\textbf {\bibinfo {volume}
  {120}},\ \bibinfo {pages} {377} (\bibinfo {year} {1987})}\BibitemShut
  {NoStop}%
\bibitem [{\citenamefont {Penrose}(1996)}]{Penrose1996}%
  \BibitemOpen
  \bibfield  {author} {\bibinfo {author} {\bibfnamefont {R.}~\bibnamefont
  {Penrose}},\ }\href {http://www.springerlink.com/index/k75046wh3668l654.pdf}
  {\bibfield  {journal} {\bibinfo  {journal} {Gen. Relativ. Gravit.}\ }\textbf
  {\bibinfo {volume} {28}},\ \bibinfo {pages} {581} (\bibinfo {year}
  {1996})}\BibitemShut {NoStop}%
\bibitem [{\citenamefont {Gasbarri}\ \emph {et~al.}(2017)\citenamefont
  {Gasbarri}, \citenamefont {Toro{\v{s}}}, \citenamefont {Donadi},\ and\
  \citenamefont {Bassi}}]{Gasbarri2017a}%
  \BibitemOpen
  \bibfield  {author} {\bibinfo {author} {\bibfnamefont {G.}~\bibnamefont
  {Gasbarri}}, \bibinfo {author} {\bibfnamefont {M.}~\bibnamefont
  {Toro{\v{s}}}}, \bibinfo {author} {\bibfnamefont {S.}~\bibnamefont {Donadi}},
  \ and\ \bibinfo {author} {\bibfnamefont {A.}~\bibnamefont {Bassi}},\ }\href
  {\doibase 10.1103/PhysRevD.96.104013} {\bibfield  {journal} {\bibinfo
  {journal} {Phys. Rev. D}\ }\textbf {\bibinfo {volume} {96}},\ \bibinfo
  {pages} {104013} (\bibinfo {year} {2017})}\BibitemShut {NoStop}%
\bibitem [{\citenamefont {Ghirardi}\ \emph
  {et~al.}(1990{\natexlab{a}})\citenamefont {Ghirardi}, \citenamefont
  {Grassi},\ and\ \citenamefont {Rimini}}]{Ghirardi1990a}%
  \BibitemOpen
  \bibfield  {author} {\bibinfo {author} {\bibfnamefont {G.}~\bibnamefont
  {Ghirardi}}, \bibinfo {author} {\bibfnamefont {R.}~\bibnamefont {Grassi}}, \
  and\ \bibinfo {author} {\bibfnamefont {A.}~\bibnamefont {Rimini}},\ }\href
  {\doibase 10.1103/PhysRevA.42.1057} {\bibfield  {journal} {\bibinfo
  {journal} {Phys. Rev. A}\ }\textbf {\bibinfo {volume} {42}},\ \bibinfo
  {pages} {1057} (\bibinfo {year} {1990}{\natexlab{a}})}\BibitemShut {NoStop}%
\bibitem [{\citenamefont {Ghirardi}\ \emph
  {et~al.}(1990{\natexlab{b}})\citenamefont {Ghirardi}, \citenamefont
  {Pearle},\ and\ \citenamefont {Rimini}}]{Ghirardi1990b}%
  \BibitemOpen
  \bibfield  {author} {\bibinfo {author} {\bibfnamefont {G.~C.}\ \bibnamefont
  {Ghirardi}}, \bibinfo {author} {\bibfnamefont {P.}~\bibnamefont {Pearle}}, \
  and\ \bibinfo {author} {\bibfnamefont {A.}~\bibnamefont {Rimini}},\ }\href
  {\doibase 10.1103/PhysRevA.42.78} {\bibfield  {journal} {\bibinfo  {journal}
  {Phys. Rev. A}\ }\textbf {\bibinfo {volume} {42}},\ \bibinfo {pages} {78}
  (\bibinfo {year} {1990}{\natexlab{b}})}\BibitemShut {NoStop}%
\bibitem [{\citenamefont {Bassi}\ and\ \citenamefont
  {Ghirardi}(2003)}]{Bassi2003}%
  \BibitemOpen
  \bibfield  {author} {\bibinfo {author} {\bibfnamefont {A.}~\bibnamefont
  {Bassi}}\ and\ \bibinfo {author} {\bibfnamefont {G.}~\bibnamefont
  {Ghirardi}},\ }\href {\doibase 10.1016/S0370-1573(03)00103-0} {\bibfield
  {journal} {\bibinfo  {journal} {Phys. Rep.}\ }\textbf {\bibinfo {volume}
  {379}},\ \bibinfo {pages} {257} (\bibinfo {year} {2003})}\BibitemShut
  {NoStop}%
\bibitem [{\citenamefont {Bassi}\ \emph {et~al.}(2013)\citenamefont {Bassi},
  \citenamefont {Lochan}, \citenamefont {Satin}, \citenamefont {Singh},\ and\
  \citenamefont {Ulbricht}}]{Bassi2012}%
  \BibitemOpen
  \bibfield  {author} {\bibinfo {author} {\bibfnamefont {A.}~\bibnamefont
  {Bassi}}, \bibinfo {author} {\bibfnamefont {K.}~\bibnamefont {Lochan}},
  \bibinfo {author} {\bibfnamefont {S.}~\bibnamefont {Satin}}, \bibinfo
  {author} {\bibfnamefont {T.~P.}\ \bibnamefont {Singh}}, \ and\ \bibinfo
  {author} {\bibfnamefont {H.}~\bibnamefont {Ulbricht}},\ }\href {\doibase
  10.1103/RevModPhys.85.471} {\bibfield  {journal} {\bibinfo  {journal} {Rev.
  Mod. Phys.}\ }\textbf {\bibinfo {volume} {85}},\ \bibinfo {pages} {471}
  (\bibinfo {year} {2013})}\BibitemShut {NoStop}%
\bibitem [{\citenamefont {Bassi}\ \emph {et~al.}(2017)\citenamefont {Bassi},
  \citenamefont {Gro{\ss}ardt},\ and\ \citenamefont {Ulbricht}}]{Bassi2017}%
  \BibitemOpen
  \bibfield  {author} {\bibinfo {author} {\bibfnamefont {A.}~\bibnamefont
  {Bassi}}, \bibinfo {author} {\bibfnamefont {A.}~\bibnamefont {Gro{\ss}ardt}},
  \ and\ \bibinfo {author} {\bibfnamefont {H.}~\bibnamefont {Ulbricht}},\
  }\href {\doibase 10.1088/1361-6382/aa864f} {\bibfield  {journal} {\bibinfo
  {journal} {Class. Quantum Grav.}\ }\textbf {\bibinfo {volume} {34}},\
  \bibinfo {pages} {193002} (\bibinfo {year} {2017})}\BibitemShut {NoStop}%
\bibitem [{\citenamefont {Ruffini}\ and\ \citenamefont
  {Bonazolla}(1969)}]{Ruffini1969}%
  \BibitemOpen
  \bibfield  {author} {\bibinfo {author} {\bibfnamefont {R.}~\bibnamefont
  {Ruffini}}\ and\ \bibinfo {author} {\bibfnamefont {S.}~\bibnamefont
  {Bonazolla}},\ }\href {\doibase 10.1103/PhysRev.187.1767} {\bibfield
  {journal} {\bibinfo  {journal} {Phys. Rev.}\ }\textbf {\bibinfo {volume}
  {187}},\ \bibinfo {pages} {1767} (\bibinfo {year} {1969})}\BibitemShut
  {NoStop}%
\bibitem [{\citenamefont {Di{\'{o}}si}(1984)}]{Diosi1984}%
  \BibitemOpen
  \bibfield  {author} {\bibinfo {author} {\bibfnamefont {L.}~\bibnamefont
  {Di{\'{o}}si}},\ }\href {\doibase 10.1016/0375-9601(84)90397-9} {\bibfield
  {journal} {\bibinfo  {journal} {Phys. Lett. A}\ }\textbf {\bibinfo {volume}
  {105}},\ \bibinfo {pages} {199} (\bibinfo {year} {1984})}\BibitemShut
  {NoStop}%
\bibitem [{\citenamefont {Carlip}(2008)}]{Carlip2008}%
  \BibitemOpen
  \bibfield  {author} {\bibinfo {author} {\bibfnamefont {S.}~\bibnamefont
  {Carlip}},\ }\href {http://iopscience.iop.org/0264-9381/25/15/154010}
  {\bibfield  {journal} {\bibinfo  {journal} {Class. Quantum Grav.}\ }\textbf
  {\bibinfo {volume} {25}},\ \bibinfo {pages} {154010} (\bibinfo {year}
  {2008})}\BibitemShut {NoStop}%
\bibitem [{\citenamefont {Giulini}\ and\ \citenamefont
  {Gro{\ss}ardt}(2011)}]{Giulini2011}%
  \BibitemOpen
  \bibfield  {author} {\bibinfo {author} {\bibfnamefont {D.}~\bibnamefont
  {Giulini}}\ and\ \bibinfo {author} {\bibfnamefont {A.}~\bibnamefont
  {Gro{\ss}ardt}},\ }\href {\doibase 10.1088/0264-9381/28/19/195026} {\bibfield
   {journal} {\bibinfo  {journal} {Class. Quantum Grav.}\ }\textbf {\bibinfo
  {volume} {28}},\ \bibinfo {pages} {195026} (\bibinfo {year}
  {2011})}\BibitemShut {NoStop}%
\bibitem [{\citenamefont {Giulini}\ and\ \citenamefont
  {Gro{\ss}ardt}(2012)}]{Giulini2012}%
  \BibitemOpen
  \bibfield  {author} {\bibinfo {author} {\bibfnamefont {D.}~\bibnamefont
  {Giulini}}\ and\ \bibinfo {author} {\bibfnamefont {A.}~\bibnamefont
  {Gro{\ss}ardt}},\ }\href {\doibase 10.1088/0264-9381/29/21/215010} {\bibfield
   {journal} {\bibinfo  {journal} {Class. Quantum Grav.}\ }\textbf {\bibinfo
  {volume} {29}},\ \bibinfo {pages} {215010} (\bibinfo {year}
  {2012})}\BibitemShut {NoStop}%
\bibitem [{\citenamefont {Bahrami}\ \emph {et~al.}(2014)\citenamefont
  {Bahrami}, \citenamefont {Gro{\ss}ardt}, \citenamefont {Donadi},\ and\
  \citenamefont {Bassi}}]{Bahrami2014d}%
  \BibitemOpen
  \bibfield  {author} {\bibinfo {author} {\bibfnamefont {M.}~\bibnamefont
  {Bahrami}}, \bibinfo {author} {\bibfnamefont {A.}~\bibnamefont
  {Gro{\ss}ardt}}, \bibinfo {author} {\bibfnamefont {S.}~\bibnamefont
  {Donadi}}, \ and\ \bibinfo {author} {\bibfnamefont {A.}~\bibnamefont
  {Bassi}},\ }\href {\doibase 10.1088/1367-2630/16/11/115007} {\bibfield
  {journal} {\bibinfo  {journal} {New J. Phys.}\ }\textbf {\bibinfo {volume}
  {16}},\ \bibinfo {pages} {115007} (\bibinfo {year} {2014})}\BibitemShut
  {NoStop}%
\bibitem [{\citenamefont {Bera}\ \emph {et~al.}(2015)\citenamefont {Bera},
  \citenamefont {Mohan},\ and\ \citenamefont {Singh}}]{Bera2015}%
  \BibitemOpen
  \bibfield  {author} {\bibinfo {author} {\bibfnamefont {S.}~\bibnamefont
  {Bera}}, \bibinfo {author} {\bibfnamefont {R.}~\bibnamefont {Mohan}}, \ and\
  \bibinfo {author} {\bibfnamefont {T.~P.}\ \bibnamefont {Singh}},\ }\href
  {\doibase 10.1103/PhysRevD.92.025054} {\bibfield  {journal} {\bibinfo
  {journal} {Phys. Rev. D}\ }\textbf {\bibinfo {volume} {92}},\ \bibinfo
  {pages} {025054} (\bibinfo {year} {2015})}\BibitemShut {NoStop}%
\bibitem [{\citenamefont {Singh}(2015)}]{Singh2015}%
  \BibitemOpen
  \bibfield  {author} {\bibinfo {author} {\bibfnamefont {T.~P.}\ \bibnamefont
  {Singh}},\ }\href {http://arxiv.org/abs/1503.01040} {\  (\bibinfo {year}
  {2015})},\ \Eprint {http://arxiv.org/abs/1503.01040} {arXiv:1503.01040}
  \BibitemShut {NoStop}%
\bibitem [{\citenamefont {Nimmrichter}\ and\ \citenamefont
  {Hornberger}(2015)}]{Nimmrichter2015}%
  \BibitemOpen
  \bibfield  {author} {\bibinfo {author} {\bibfnamefont {S.}~\bibnamefont
  {Nimmrichter}}\ and\ \bibinfo {author} {\bibfnamefont {K.}~\bibnamefont
  {Hornberger}},\ }\href {\doibase 10.1103/PhysRevD.91.024016} {\bibfield
  {journal} {\bibinfo  {journal} {Phys. Rev. D}\ }\textbf {\bibinfo {volume}
  {91}},\ \bibinfo {pages} {024016} (\bibinfo {year} {2015})}\BibitemShut
  {NoStop}%
\bibitem [{\citenamefont {Bera}\ \emph {et~al.}(2017)\citenamefont {Bera},
  \citenamefont {Giri},\ and\ \citenamefont {Singh}}]{Bera2017}%
  \BibitemOpen
  \bibfield  {author} {\bibinfo {author} {\bibfnamefont {S.}~\bibnamefont
  {Bera}}, \bibinfo {author} {\bibfnamefont {P.}~\bibnamefont {Giri}}, \ and\
  \bibinfo {author} {\bibfnamefont {T.~P.}\ \bibnamefont {Singh}},\ }\href
  {\doibase 10.1007/s10701-017-0092-5} {\bibfield  {journal} {\bibinfo
  {journal} {Found. Phys.}\ }\textbf {\bibinfo {volume} {47}},\ \bibinfo
  {pages} {897} (\bibinfo {year} {2017})}\BibitemShut {NoStop}%
\bibitem [{\citenamefont {Kafri}\ \emph {et~al.}(2014)\citenamefont {Kafri},
  \citenamefont {Taylor},\ and\ \citenamefont {Milburn}}]{Kafri2014}%
  \BibitemOpen
  \bibfield  {author} {\bibinfo {author} {\bibfnamefont {D.}~\bibnamefont
  {Kafri}}, \bibinfo {author} {\bibfnamefont {J.~M.}\ \bibnamefont {Taylor}}, \
  and\ \bibinfo {author} {\bibfnamefont {G.~J.}\ \bibnamefont {Milburn}},\
  }\href {\doibase 10.1088/1367-2630/16/6/065020} {\bibfield  {journal}
  {\bibinfo  {journal} {New J. Phys.}\ }\textbf {\bibinfo {volume} {16}},\
  \bibinfo {pages} {065020} (\bibinfo {year} {2014})}\BibitemShut {NoStop}%
\bibitem [{\citenamefont {Kafri}\ \emph {et~al.}(2015)\citenamefont {Kafri},
  \citenamefont {Milburn},\ and\ \citenamefont {Taylor}}]{Kafri2015}%
  \BibitemOpen
  \bibfield  {author} {\bibinfo {author} {\bibfnamefont {D.}~\bibnamefont
  {Kafri}}, \bibinfo {author} {\bibfnamefont {G.~J.}\ \bibnamefont {Milburn}},
  \ and\ \bibinfo {author} {\bibfnamefont {J.~M.}\ \bibnamefont {Taylor}},\
  }\href {\doibase 10.1088/1367-2630/17/1/015006} {\bibfield  {journal}
  {\bibinfo  {journal} {New J. Phys.}\ }\textbf {\bibinfo {volume} {17}},\
  \bibinfo {pages} {015006} (\bibinfo {year} {2015})}\BibitemShut {NoStop}%
\bibitem [{\citenamefont {Tilloy}\ and\ \citenamefont
  {Di{\'{o}}si}(2016)}]{Tilloy2016}%
  \BibitemOpen
  \bibfield  {author} {\bibinfo {author} {\bibfnamefont {A.}~\bibnamefont
  {Tilloy}}\ and\ \bibinfo {author} {\bibfnamefont {L.}~\bibnamefont
  {Di{\'{o}}si}},\ }\href {\doibase 10.1103/PhysRevD.93.024026} {\bibfield
  {journal} {\bibinfo  {journal} {Phys. Rev. D}\ }\textbf {\bibinfo {volume}
  {93}},\ \bibinfo {pages} {024026} (\bibinfo {year} {2016})}\BibitemShut
  {NoStop}%
\bibitem [{\citenamefont {Tilloy}(2018)}]{Tilloy2018}%
  \BibitemOpen
  \bibfield  {author} {\bibinfo {author} {\bibfnamefont {A.}~\bibnamefont
  {Tilloy}},\ }\href {\doibase 10.1103/PhysRevD.97.021502} {\bibfield
  {journal} {\bibinfo  {journal} {Phys. Rev. D}\ }\textbf {\bibinfo {volume}
  {97}},\ \bibinfo {pages} {021502} (\bibinfo {year} {2018})}\BibitemShut
  {NoStop}%
\bibitem [{\citenamefont {Leggett}(2002)}]{Leggett2002}%
  \BibitemOpen
  \bibfield  {author} {\bibinfo {author} {\bibfnamefont {A.~J.}\ \bibnamefont
  {Leggett}},\ }\href {\doibase 10.1088/0953-8984/14/15/201} {\bibfield
  {journal} {\bibinfo  {journal} {J. Phys. Condens. Matter}\ }\textbf {\bibinfo
  {volume} {14}},\ \bibinfo {pages} {R415} (\bibinfo {year}
  {2002})}\BibitemShut {NoStop}%
\bibitem [{\citenamefont {Nimmrichter}\ and\ \citenamefont
  {Hornberger}(2013)}]{Nimmrichter2013}%
  \BibitemOpen
  \bibfield  {author} {\bibinfo {author} {\bibfnamefont {S.}~\bibnamefont
  {Nimmrichter}}\ and\ \bibinfo {author} {\bibfnamefont {K.}~\bibnamefont
  {Hornberger}},\ }\href {\doibase 10.1103/PhysRevLett.110.160403} {\bibfield
  {journal} {\bibinfo  {journal} {Phys. Rev. Lett.}\ }\textbf {\bibinfo
  {volume} {110}},\ \bibinfo {pages} {160403} (\bibinfo {year}
  {2013})}\BibitemShut {NoStop}%
\bibitem [{\citenamefont {Wiseman}\ and\ \citenamefont
  {Milburn}(2010)}]{Wiseman2010}%
  \BibitemOpen
  \bibfield  {author} {\bibinfo {author} {\bibfnamefont {H.~M.}\ \bibnamefont
  {Wiseman}}\ and\ \bibinfo {author} {\bibfnamefont {G.~J.}\ \bibnamefont
  {Milburn}},\ }\href {\doibase 10.1017/CBO9780511813948} {\emph {\bibinfo
  {title} {{Quantum measurement and control}}}}\ (\bibinfo  {publisher}
  {Cambridge University Press},\ \bibinfo {year} {2010})\BibitemShut {NoStop}%
\bibitem [{\citenamefont {Jacobs}(2014)}]{Jacobs2014}%
  \BibitemOpen
  \bibfield  {author} {\bibinfo {author} {\bibfnamefont {K.}~\bibnamefont
  {Jacobs}},\ }\href {\doibase 10.1017/CBO9781139179027} {\emph {\bibinfo
  {title} {{Quantum Measurement Theory and its Applications}}}}\ (\bibinfo
  {publisher} {Cambridge University Press},\ \bibinfo {address} {Cambridge},\
  \bibinfo {year} {2014})\BibitemShut {NoStop}%
\bibitem [{\citenamefont {Adler}(2004)}]{Adler2004}%
  \BibitemOpen
  \bibfield  {author} {\bibinfo {author} {\bibfnamefont {S.~L.}\ \bibnamefont
  {Adler}},\ }\href
  {http://www.amazon.com/Quantum-Theory-Emergent-Phenomenon-Statistical/dp/0521831946}
  {\emph {\bibinfo {title} {{Quantum Theory as an Emergent Phenomenon: The
  Statistical Mechanics of Matrix Models as the Precursor of Quantum Field
  Theory}}}}\ (\bibinfo  {publisher} {Cambridge University Press},\ \bibinfo
  {address} {Cambridge},\ \bibinfo {year} {2004})\BibitemShut {NoStop}%
\bibitem [{\citenamefont {Ghirardi}\ \emph {et~al.}(1986)\citenamefont
  {Ghirardi}, \citenamefont {Rimini},\ and\ \citenamefont
  {Weber}}]{Ghirardi1986}%
  \BibitemOpen
  \bibfield  {author} {\bibinfo {author} {\bibfnamefont {G.~C.}\ \bibnamefont
  {Ghirardi}}, \bibinfo {author} {\bibfnamefont {A.}~\bibnamefont {Rimini}}, \
  and\ \bibinfo {author} {\bibfnamefont {T.}~\bibnamefont {Weber}},\ }\href
  {\doibase 10.1103/PhysRevD.34.470} {\bibfield  {journal} {\bibinfo  {journal}
  {Phys. Rev. D}\ }\textbf {\bibinfo {volume} {34}},\ \bibinfo {pages} {470}
  (\bibinfo {year} {1986})}\BibitemShut {NoStop}%
\bibitem [{\citenamefont {Vacchini}(2007)}]{Vacchini2007b}%
  \BibitemOpen
  \bibfield  {author} {\bibinfo {author} {\bibfnamefont {B.}~\bibnamefont
  {Vacchini}},\ }\href {\doibase 10.1088/1751-8113/40/10/015} {\bibfield
  {journal} {\bibinfo  {journal} {J. Phys. A}\ }\textbf {\bibinfo {volume}
  {40}},\ \bibinfo {pages} {2463} (\bibinfo {year} {2007})}\BibitemShut
  {NoStop}%
\bibitem [{\citenamefont {Khosla}\ and\ \citenamefont
  {Altamirano}(2017)}]{Khosla2017}%
  \BibitemOpen
  \bibfield  {author} {\bibinfo {author} {\bibfnamefont {K.~E.}\ \bibnamefont
  {Khosla}}\ and\ \bibinfo {author} {\bibfnamefont {N.}~\bibnamefont
  {Altamirano}},\ }\href {\doibase 10.1103/PhysRevA.95.052116} {\bibfield
  {journal} {\bibinfo  {journal} {Phys. Rev. A}\ }\textbf {\bibinfo {volume}
  {95}},\ \bibinfo {pages} {052116} (\bibinfo {year} {2017})}\BibitemShut
  {NoStop}%
\bibitem [{\citenamefont {Altamirano}\ \emph {et~al.}(2018)\citenamefont
  {Altamirano}, \citenamefont {Corona-Ugalde}, \citenamefont {Mann},\ and\
  \citenamefont {Zych}}]{Altamirano2018}%
  \BibitemOpen
  \bibfield  {author} {\bibinfo {author} {\bibfnamefont {N.}~\bibnamefont
  {Altamirano}}, \bibinfo {author} {\bibfnamefont {P.}~\bibnamefont
  {Corona-Ugalde}}, \bibinfo {author} {\bibfnamefont {R.~B.}\ \bibnamefont
  {Mann}}, \ and\ \bibinfo {author} {\bibfnamefont {M.}~\bibnamefont {Zych}},\
  }\href {\doibase 10.1088/1361-6382/aac72f} {\bibfield  {journal} {\bibinfo
  {journal} {Class. Quantum Grav.}\ }\textbf {\bibinfo {volume} {35}},\
  \bibinfo {pages} {145005} (\bibinfo {year} {2018})}\BibitemShut {NoStop}%
\bibitem [{\citenamefont {Peters}\ \emph {et~al.}(2001)\citenamefont {Peters},
  \citenamefont {Chung},\ and\ \citenamefont {Chu}}]{Peters2001}%
  \BibitemOpen
  \bibfield  {author} {\bibinfo {author} {\bibfnamefont {A.}~\bibnamefont
  {Peters}}, \bibinfo {author} {\bibfnamefont {K.}~\bibnamefont {Chung}}, \
  and\ \bibinfo {author} {\bibfnamefont {S.}~\bibnamefont {Chu}},\ }\href
  {http://iopscience.iop.org/0026-1394/38/1/4} {\bibfield  {journal} {\bibinfo
  {journal} {Metrologia}\ }\textbf {\bibinfo {volume} {38}},\ \bibinfo {pages}
  {25} (\bibinfo {year} {2001})}\BibitemShut {NoStop}%
\bibitem [{\citenamefont {Rosi}\ \emph {et~al.}(2014)\citenamefont {Rosi},
  \citenamefont {Sorrentino}, \citenamefont {Cacciapuoti}, \citenamefont
  {Prevedelli},\ and\ \citenamefont {Tino}}]{Rosi2014}%
  \BibitemOpen
  \bibfield  {author} {\bibinfo {author} {\bibfnamefont {G.}~\bibnamefont
  {Rosi}}, \bibinfo {author} {\bibfnamefont {F.}~\bibnamefont {Sorrentino}},
  \bibinfo {author} {\bibfnamefont {L.}~\bibnamefont {Cacciapuoti}}, \bibinfo
  {author} {\bibfnamefont {M.}~\bibnamefont {Prevedelli}}, \ and\ \bibinfo
  {author} {\bibfnamefont {G.~M.}\ \bibnamefont {Tino}},\ }\href {\doibase
  10.1038/nature13433} {\bibfield  {journal} {\bibinfo  {journal} {Nature}\
  }\textbf {\bibinfo {volume} {510}},\ \bibinfo {pages} {518} (\bibinfo {year}
  {2014})}\BibitemShut {NoStop}%
\bibitem [{\citenamefont {Biedermann}\ \emph {et~al.}(2015)\citenamefont
  {Biedermann}, \citenamefont {Wu}, \citenamefont {Deslauriers}, \citenamefont
  {Roy}, \citenamefont {Mahadeswaraswamy},\ and\ \citenamefont
  {Kasevich}}]{Biedermann2015}%
  \BibitemOpen
  \bibfield  {author} {\bibinfo {author} {\bibfnamefont {G.~W.}\ \bibnamefont
  {Biedermann}}, \bibinfo {author} {\bibfnamefont {X.}~\bibnamefont {Wu}},
  \bibinfo {author} {\bibfnamefont {L.}~\bibnamefont {Deslauriers}}, \bibinfo
  {author} {\bibfnamefont {S.}~\bibnamefont {Roy}}, \bibinfo {author}
  {\bibfnamefont {C.}~\bibnamefont {Mahadeswaraswamy}}, \ and\ \bibinfo
  {author} {\bibfnamefont {M.~A.}\ \bibnamefont {Kasevich}},\ }\href {\doibase
  10.1103/PhysRevA.91.033629} {\bibfield  {journal} {\bibinfo  {journal} {Phys.
  Rev. A}\ }\textbf {\bibinfo {volume} {91}},\ \bibinfo {pages} {033629}
  (\bibinfo {year} {2015})}\BibitemShut {NoStop}%
\bibitem [{\citenamefont {Hamilton}\ \emph {et~al.}(2015)\citenamefont
  {Hamilton}, \citenamefont {Jaffe}, \citenamefont {Haslinger}, \citenamefont
  {Simmons}, \citenamefont {M{\"{u}}ller},\ and\ \citenamefont
  {Khoury}}]{Hamilton2015}%
  \BibitemOpen
  \bibfield  {author} {\bibinfo {author} {\bibfnamefont {P.}~\bibnamefont
  {Hamilton}}, \bibinfo {author} {\bibfnamefont {M.}~\bibnamefont {Jaffe}},
  \bibinfo {author} {\bibfnamefont {P.}~\bibnamefont {Haslinger}}, \bibinfo
  {author} {\bibfnamefont {Q.}~\bibnamefont {Simmons}}, \bibinfo {author}
  {\bibfnamefont {H.}~\bibnamefont {M{\"{u}}ller}}, \ and\ \bibinfo {author}
  {\bibfnamefont {J.}~\bibnamefont {Khoury}},\ }\href {\doibase
  10.1126/science.aaa8883} {\bibfield  {journal} {\bibinfo  {journal}
  {Science}\ }\textbf {\bibinfo {volume} {349}},\ \bibinfo {pages} {849}
  (\bibinfo {year} {2015})}\BibitemShut {NoStop}%
\bibitem [{\citenamefont {Kovachy}\ \emph {et~al.}(2015)\citenamefont
  {Kovachy}, \citenamefont {Asenbaum}, \citenamefont {Overstreet},
  \citenamefont {Donnelly}, \citenamefont {Dickerson}, \citenamefont
  {Sugarbaker}, \citenamefont {Hogan},\ and\ \citenamefont
  {Kasevich}}]{Kovachy2015a}%
  \BibitemOpen
  \bibfield  {author} {\bibinfo {author} {\bibfnamefont {T.}~\bibnamefont
  {Kovachy}}, \bibinfo {author} {\bibfnamefont {P.}~\bibnamefont {Asenbaum}},
  \bibinfo {author} {\bibfnamefont {C.}~\bibnamefont {Overstreet}}, \bibinfo
  {author} {\bibfnamefont {C.~A.}\ \bibnamefont {Donnelly}}, \bibinfo {author}
  {\bibfnamefont {S.~M.}\ \bibnamefont {Dickerson}}, \bibinfo {author}
  {\bibfnamefont {A.}~\bibnamefont {Sugarbaker}}, \bibinfo {author}
  {\bibfnamefont {J.~M.}\ \bibnamefont {Hogan}}, \ and\ \bibinfo {author}
  {\bibfnamefont {M.~A.}\ \bibnamefont {Kasevich}},\ }\href {\doibase
  10.1038/nature16155} {\bibfield  {journal} {\bibinfo  {journal} {Nature}\
  }\textbf {\bibinfo {volume} {528}},\ \bibinfo {pages} {530} (\bibinfo {year}
  {2015})}\BibitemShut {NoStop}%
\bibitem [{\citenamefont {Jaffe}\ \emph {et~al.}(2017)\citenamefont {Jaffe},
  \citenamefont {Haslinger}, \citenamefont {Xu}, \citenamefont {Hamilton},
  \citenamefont {Upadhye}, \citenamefont {Elder}, \citenamefont {Khoury},\ and\
  \citenamefont {M{\"{u}}ller}}]{Jaffe2017}%
  \BibitemOpen
  \bibfield  {author} {\bibinfo {author} {\bibfnamefont {M.}~\bibnamefont
  {Jaffe}}, \bibinfo {author} {\bibfnamefont {P.}~\bibnamefont {Haslinger}},
  \bibinfo {author} {\bibfnamefont {V.}~\bibnamefont {Xu}}, \bibinfo {author}
  {\bibfnamefont {P.}~\bibnamefont {Hamilton}}, \bibinfo {author}
  {\bibfnamefont {A.}~\bibnamefont {Upadhye}}, \bibinfo {author} {\bibfnamefont
  {B.}~\bibnamefont {Elder}}, \bibinfo {author} {\bibfnamefont
  {J.}~\bibnamefont {Khoury}}, \ and\ \bibinfo {author} {\bibfnamefont
  {H.}~\bibnamefont {M{\"{u}}ller}},\ }\href {\doibase 10.1038/nphys4189}
  {\bibfield  {journal} {\bibinfo  {journal} {Nat. Phys.}\ }\textbf {\bibinfo
  {volume} {13}},\ \bibinfo {pages} {938} (\bibinfo {year} {2017})}\BibitemShut
  {NoStop}%
\bibitem [{\citenamefont {Nesvizhevsky}\ \emph {et~al.}(2002)\citenamefont
  {Nesvizhevsky}, \citenamefont {B{\"{o}}rner}, \citenamefont {Petukhov},
  \citenamefont {Abele}, \citenamefont {Bae{\ss}ler}, \citenamefont {Rue{\ss}},
  \citenamefont {St{\"{o}}ferle}, \citenamefont {Westphal}, \citenamefont
  {Gagarski}, \citenamefont {Petrov},\ and\ \citenamefont
  {Strelkov}}]{Nesvizhevsky2002}%
  \BibitemOpen
  \bibfield  {author} {\bibinfo {author} {\bibfnamefont {V.~V.}\ \bibnamefont
  {Nesvizhevsky}}, \bibinfo {author} {\bibfnamefont {H.~G.}\ \bibnamefont
  {B{\"{o}}rner}}, \bibinfo {author} {\bibfnamefont {A.~K.}\ \bibnamefont
  {Petukhov}}, \bibinfo {author} {\bibfnamefont {H.}~\bibnamefont {Abele}},
  \bibinfo {author} {\bibfnamefont {S.}~\bibnamefont {Bae{\ss}ler}}, \bibinfo
  {author} {\bibfnamefont {F.~J.}\ \bibnamefont {Rue{\ss}}}, \bibinfo {author}
  {\bibfnamefont {T.}~\bibnamefont {St{\"{o}}ferle}}, \bibinfo {author}
  {\bibfnamefont {A.}~\bibnamefont {Westphal}}, \bibinfo {author}
  {\bibfnamefont {A.~M.}\ \bibnamefont {Gagarski}}, \bibinfo {author}
  {\bibfnamefont {G.~A.}\ \bibnamefont {Petrov}}, \ and\ \bibinfo {author}
  {\bibfnamefont {A.~V.}\ \bibnamefont {Strelkov}},\ }\href {\doibase
  10.1038/415297a} {\bibfield  {journal} {\bibinfo  {journal} {Nature}\
  }\textbf {\bibinfo {volume} {415}},\ \bibinfo {pages} {297} (\bibinfo {year}
  {2002})}\BibitemShut {NoStop}%
\bibitem [{\citenamefont {Jenke}\ \emph {et~al.}(2011)\citenamefont {Jenke},
  \citenamefont {Geltenbort}, \citenamefont {Lemmel},\ and\ \citenamefont
  {Abele}}]{Jenke2011}%
  \BibitemOpen
  \bibfield  {author} {\bibinfo {author} {\bibfnamefont {T.}~\bibnamefont
  {Jenke}}, \bibinfo {author} {\bibfnamefont {P.}~\bibnamefont {Geltenbort}},
  \bibinfo {author} {\bibfnamefont {H.}~\bibnamefont {Lemmel}}, \ and\ \bibinfo
  {author} {\bibfnamefont {H.}~\bibnamefont {Abele}},\ }\href {\doibase
  10.1038/nphys1970} {\bibfield  {journal} {\bibinfo  {journal} {Nat. Phys.}\
  }\textbf {\bibinfo {volume} {7}},\ \bibinfo {pages} {468} (\bibinfo {year}
  {2011})}\BibitemShut {NoStop}%
\bibitem [{\citenamefont {Abele}\ and\ \citenamefont {Leeb}(2012)}]{Abele2012}%
  \BibitemOpen
  \bibfield  {author} {\bibinfo {author} {\bibfnamefont {H.}~\bibnamefont
  {Abele}}\ and\ \bibinfo {author} {\bibfnamefont {H.}~\bibnamefont {Leeb}},\
  }\href {\doibase 10.1088/1367-2630/14/5/055010} {\bibfield  {journal}
  {\bibinfo  {journal} {New J. Phys.}\ }\textbf {\bibinfo {volume} {14}},\
  \bibinfo {pages} {055010} (\bibinfo {year} {2012})}\BibitemShut {NoStop}%
\bibitem [{\citenamefont {Cronenberg}\ \emph {et~al.}(2015)\citenamefont
  {Cronenberg}, \citenamefont {Filter}, \citenamefont {Thalhammer},
  \citenamefont {Jenke}, \citenamefont {Abele},\ and\ \citenamefont
  {Geltenbort}}]{Cronenberg2015}%
  \BibitemOpen
  \bibfield  {author} {\bibinfo {author} {\bibfnamefont {G.}~\bibnamefont
  {Cronenberg}}, \bibinfo {author} {\bibfnamefont {H.}~\bibnamefont {Filter}},
  \bibinfo {author} {\bibfnamefont {M.}~\bibnamefont {Thalhammer}}, \bibinfo
  {author} {\bibfnamefont {T.}~\bibnamefont {Jenke}}, \bibinfo {author}
  {\bibfnamefont {H.}~\bibnamefont {Abele}}, \ and\ \bibinfo {author}
  {\bibfnamefont {P.}~\bibnamefont {Geltenbort}},\ }\href
  {http://arxiv.org/abs/1512.09134} {\  (\bibinfo {year} {2015})},\ \Eprint
  {http://arxiv.org/abs/1512.09134} {arXiv:1512.09134} \BibitemShut {NoStop}%
\end{thebibliography}%

\end{document}